\documentclass[final,5p,times,twocolumn]{elsarticleRN}

\usepackage{graphics,amsmath}

\usepackage{amssymb,url}

\journal{Journal of Crystal Growth}

\newcommand{\bR}{\mathbb{R}}
\newcommand{\Rgeq}{\bR_{\geq 0}}
\newcommand{\ds}{\;{\rm d}s} 
\newcommand{\uD}{u_{\partial\Omega}}              
\newcommand{\dd}[1]{\frac{\rm d}{{\rm d}#1}}
\newcommand{\ddt}{\dd{t}}

\begin{document}

\begin{frontmatter}



\title{Numerical computations of facetted pattern formation in snow
  crystal growth}

\author[label1]{John W.\ Barrett}
\address[label1]{Department of Mathematics, Imperial College London, London SW7 2AZ, UK}
\author{Harald Garcke\corref{cor1}\fnref{label2}}
\address[label2]{Fakult\"at f\"ur Mathematik, Universit\"at Regensburg, 93040
 Regensburg, Germany}
 \cortext[cor1]{Corresponding author. Tel.: +49 941 943 2992; fax: +
     49 941 943 3263}
\ead{harald.garcke@mathematik.uni-regensburg.de}
\author[label1]{Robert N\"urnberg}

\begin{abstract}
Facetted growth of snow crystals leads to a rich diversity of forms, and
exhibits a remarkable sixfold symmetry. Snow crystal structures result
from diffusion limited crystal growth in the presence of anisotropic
surface energy and anisotropic attachment kinetics. 
It is by now well understood 
that the morphological stability of ice crystals strongly
depends on supersaturation, crystal size and temperature. 
Until very recently it was very
difficult to perform numerical simulations of this highly anisotropic
crystal growth. In particular, obtaining facet growth in combination
with dendritic branching is a challenging task. We present numerical
simulations of snow crystal growth in two and three space dimensions 
using a new computational method
recently introduced by the authors. We present both qualitative and
quantitative computations. In particular, a linear relationship between tip
velocity and supersaturation is observed. The computations also
suggest that surface energy effects, although small, have a larger
effect on crystal growth than previously expected. We compute
solid plates, solid prisms, hollow columns, needles, dendrites, capped
columns and scrolls on plates. Although all these forms appear in
nature, most of these forms are computed here for the first time in
numerical simulations for a continuum model.
\end{abstract}


\end{frontmatter}



\section{Introduction}
\label{intro}

\begin{figure}
\center
\includegraphics[angle=0,height=6cm]{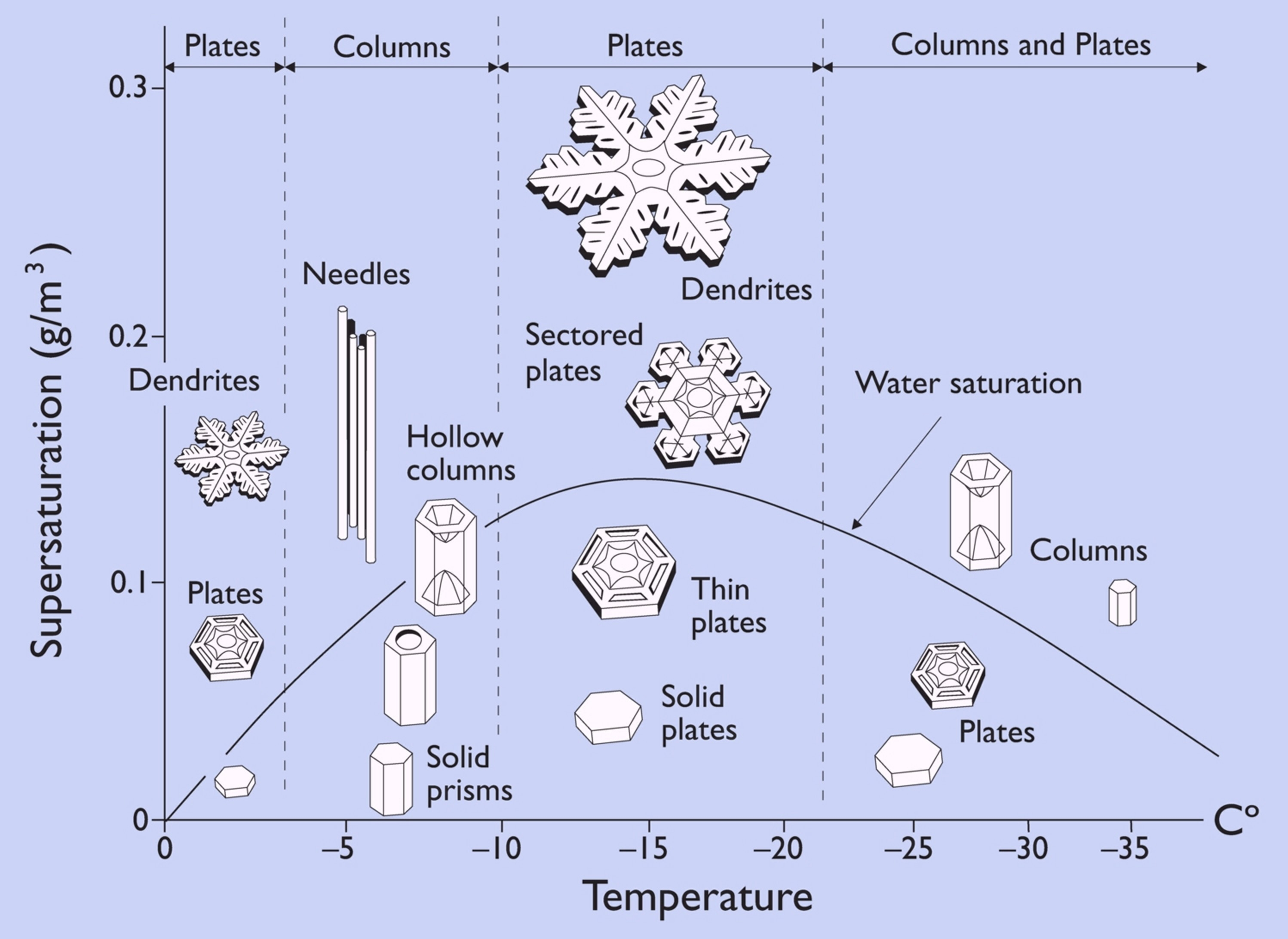}
\caption{The Nakaya diagram illustrates which snow crystal forms
  appear at different temperatures and supersaturations. This figure
  is taken from \cite{Libbrecht05}.}
\label{fig:libbrecht} 
\end{figure}

Snow crystals grown from a supersaturated vapour lead to a variety of
complex and often very symmetric patterns. Crystallisation from vapour
is a fundamental phase transition, and a good understanding is crucial
for many applications. Numerous experiments have been performed,
and compilations of photographs of artificial and natural snowflakes
reveal their beauty and complexity, see \cite{Nakaya54} and the review
\cite{Libbrecht05}. The precise forms of snow crystals depend in a very
subtle way on the temperature and the supersaturation. \citet{Nakaya54} 
analysed these dependencies in detail and combined his observations in his now
famous Nakaya {\it snow crystal morphology diagram}, see
Figure~\ref{fig:libbrecht}. At temperatures just below the freezing
temperature thick plates grow at lower supersaturations and plate-like
dendritic forms appear at higher supersaturations. At temperatures
around $-5^\circ$C solid prisms grow at lower supersaturations and
hollow columns and needle-like crystals at higher supersaturations. If
the temperature is decreased below $-10^\circ$C one observes thin
solid plates at low supersaturations, whereas dendrites form at
high supersaturations. Below $-25^\circ$C again columns form at high
supersaturations. The results from \citet{Nakaya54}, 
which led to the snow crystal morphology diagram, 
have been confirmed by many subsequent experimental
studies. Although the experiments give a rather clear picture,
the physics behind the snow crystal morphology
diagram are not yet understood. 

A continuum mathematical modelling of snow crystal growth leads to a
quasi-static diffusion problem for the diffusion of the vapour
molecules. The diffusion equation has to be solved together with
rather complex boundary conditions on the free boundary between vapour
and solid. The conditions on this interface are given by the continuity
equation relating the flux of vapour molecules onto the interface to
the interface velocity, and an equation describing the attachment
kinetics taking surface energy effects into account. In the latter
condition the hexagonal anisotropy of snow crystals also enters. Due
to being highly nonlinear, and since it is geometrically very involved,
the complete free boundary problem is difficult to analyse
theoretically. However, there exists a large literature on numerical 
computations for diffusion limited growth and the formation of
dendrites, which we now briefly discuss. 

Numerical approaches for crystal growth usually employ either sharp interface
models, in which the solidification front is tracked explicitly,
or phase field models, in which the solidification front is modelled by a
thin diffusional layer. 
In sharp interface approaches the front is
described with the help of a parameterisation, see 
\cite{RoosenT91,Yokoyama93,Schmidt96,dendritic}, or by using a level set
function, see \cite{SethianS92}. In a phase field method a new
order parameter --the phase field-- is introduced, which at the
interface rapidly changes its value between two fixed values, which
describe the different phases, 
see \cite{Kobayashi93,WheelerMS93,KarmaR98,DebierKCG03}. 
A popular discrete model for the simulation of crystal growth are 
cellular automata, see \cite{Reiter05,Libbrecht08,GravnerG09}.
Moreover, molecular dynamics
simulations are used to understand the surface structure of ice
\cite{FurukawaN97}. Although many computations
have been performed, a quantitative numerical description of facet
growth in combination with dendritic branching is missing. 

In recent research by the authors a new parametric approach for
interface motion has been developed \cite{triplej,gflows3d}. The
method has the feature that the mesh quality of the interface
approximation, which is given by a polyhedral surface mesh, 
remains good during the evolution -- 
most earlier approaches had to deal with mesh degeneracies, e.g.\ by re-meshing
the interface approximation.
In addition, the present authors were able to include anisotropy effects
into curvature driven hypersurface evolution in a numerically stable
way. This allows the method to compute in situations in which the
anisotropy is facetted, see \cite{triplejANI,ani3d,ejam3d,dendritic}.

It is the goal of this paper to demonstrate that with the numerical method
introduced in \cite{triplej,gflows3d,ani3d,dendritic} it is possible to
compute snow crystal growth in a qualitatively and quantitatively
satisfactory way. We will present computations showing a significant
number of different types of snow crystals, such as {\it solid plates},
{\it solid prisms}, {\it hollow columns}, {\it needles}, {\it
  dendrites}, {\it capped columns} and {\it scrolls on plates}. 

In snow crystal growth models several parameters are not
known to a sufficient precision. In particular, the surface energy
density as a function of orientation is not known in detail. In
addition, the condensation coefficient, which embodies the attachment
kinetics of how water molecules are incorporated into the ice lattice, is
not known in detail.
For example, it is not known how the condensation
coefficient depends on the crystal orientation, see \cite{Libbrecht05}
and the references therein for details. 

Our numerical computations seem to suggest that the anisotropy in the
surface energy density might have a more important impact on snow
crystal growth morphologies than previously expected. Of course, numerical
simulations alone cannot decide whether this is in fact the case, but a
comparison of numerical computations with experiments might help to
understand this issue better. In particular, it might be
possible to obtain more precise estimates for the size of the
condensation coefficient as a function of orientation.

We also present some quantitative results, which first of all show what
the relatives sizes of the quantities entering the surface attachment
kinetics are. In addition, we show that the tip velocity for a dendrite
growing into a supersaturated vapour depends linearly on the
supersaturation. This linear relationship has been already 
observed in experiments for growing needles, see
\cite{LibbrechtCS02}, and our computations might help to relate parameters
in the theoretical model to experiments. 

\section{A continuum model for snow crystal growth}\label{sec2}

We consider a
continuum model for snow crystal growth, 
which consists of an 
ice crystal growing from water vapour,
as discussed in e.g.\ \cite{Libbrecht05,BenJacob93},
and non-dimensionalize it. Let $c$ denote the water
vapour number density in gas. The diffusion equation in
the gas phase, see \cite[Eq.~(2)]{Libbrecht05}, is then 
\begin{equation} \label{eq:nondima}
c_t - \mathcal{D}\,\Delta\, c = 0 \,\,\mbox{ in } \,\,\Omega_+(t)\,,
\end{equation}
where $\Omega_+(t)$ is the domain occupied by the gas phase and
$\mathcal{D}$ is the corresponding diffusion constant. The mass
balance at the gas/solid interface $\Gamma(t)$ gives rise to 
\begin{equation}
\mathcal{D}\, \frac{\partial  c}{\partial\vec\nu} 
=(c_{\rm solid}-c)\,\mathcal{V} 
\,\,\mbox{ on } \,\,\Gamma(t)\,, 
 \end{equation}
where $c_{\rm solid} \approx 3\times 10^{28}\,{\rm m}^{-3}$ 
is the number density for ice. In addition, $\vec{\nu}$ is the unit
normal to $\Gamma(t)$ pointing into $\Omega_+(t)$ and $\mathcal{V}$ is
the velocity of $\Gamma(t)$ in the direction $\vec\nu$. 
In \cite[Eq.~(3)]{Libbrecht05} the term $c\,\mathcal{V}$ is neglected since
$c \ll c_{\rm solid}$. 
Furthermore, taking surface tension effects and attachment kinetics
into account, we require, compare \cite[Eq.~(23)]{Libbrecht05},
$$
c = c_{\rm sat} \,(1-\delta\,\kappa_\gamma +
\frac{\mathcal{V}}{\beta(\vec{\nu})\,v_{\rm kin}}) \,\,\mbox{ on } \,\,
\Gamma(t)\,.
$$
Here $v_{\rm kin}$ is the kinetic velocity, $\delta =
\hat{\gamma}/ (c_{\rm solid}\,K\,T)\approx 1\,{\rm nm} = 
10^{-3}\mu{\rm m}$, where $\hat{\gamma} \approx 0.1\, \rm{ J m}^{-2}$
represents the typical order of the surface tension of ice, 
$K \approx 1.4 \times 10^{-23}\,{\rm JK}^{-1}$ is
the Boltzmann constant, $T$ is the temperature
and $c_{\rm sat} = c_{\rm sat}(T)$ is the equilibrium number density above 
a flat ice surface, which is dependent on temperature. In addition,
$\kappa_\gamma$ is the anisotropic mean curvature which incorporates
the hexagonal anisotropy of the surface energy density.
Moreover, $\beta$ is the condensation coefficient, which is denoted by 
$\alpha$ in \cite{Libbrecht05}, which depends on the orientation of
the crystal via the normal $\vec{\nu}$.   
Finally, we complement (\ref{eq:nondima}) with the boundary condition
\begin{equation} \label{eq:nondimb}
c = c_\infty \,\, \mbox{ on } \,\, \partial\Omega
= \partial\Omega_+(t) \setminus \Gamma(t)\,,
\end{equation}
where $c_\infty := c_{\rm sat} + c_{\rm super}$ 
describes the water vapour number density far away
from the interface. Here, for convenience, we choose a domain
$\Omega\subset \bR^d$, $d=2,3$, with $\Omega_+(t) \subset \Omega$,
that is large enough so that boundary effects can be neglected.
Moreover,
$c_{\rm super}$ is related to the supersaturation $\varrho_{\rm super}$ by
\begin{equation}\label{eq:super}
\varrho_{\rm super} = {\rm m}_{{\rm H_2O}}\,c_{\rm super}\,,
\end{equation}
with ${\rm m}_{\rm H_2O} \approx 3\,\times10^{-23}\,{\rm g}$ denoting the mass 
of a water molecule. We recall that the supersaturation $\varrho_{\rm super}$ 
appears on the vertical axis in Figure~\ref{fig:libbrecht}.

It remains to introduce the anisotropic mean curvature
$\kappa_\gamma$. Instead of a constant surface energy density, we 
choose $\gamma$ to be dependent on the orientation of the interface.
The effect of the underlying crystal structure is
encoded into the surface energy by allowing $\gamma=\gamma (\vec{\nu})$,
where as stated above, $\vec{\nu}$ is the unit normal to the solid boundary
$\Gamma(t)$ pointing into the vapour region $\Omega_+(t)$. 
The total surface energy of an
interface $\Gamma$ is now given by the surface integral
$$
\int_\Gamma \gamma(\vec{\nu}) \ds\,.
$$
It is convenient to extend $\gamma$ to be a positively
homogeneous function of degree one, i.e.\ we define
$\tilde\gamma(\vec p) = |\vec p|\,\gamma(\vec p / |\vec p|)$ for all
$\vec p \not=\vec 0$ and refer to $\tilde\gamma$ as $\gamma$ from now on. 
The first variation of the above energy can
now be computed as 
$$
\kappa_\gamma := -\nabla_s\cdot\gamma'(\vec{\nu})\,,
$$
i.e.\ $\ddt \int_{\Gamma(t)} \gamma(\vec{\nu}) \ds =
- \int_{\Gamma(t)} \kappa_\gamma\,\mathcal{V}\ds$;
where $\nabla_s\,\cdot$ is the tangential divergence on $\Gamma$ and
$\gamma'$ is the gradient of $\gamma$, see e.g.\ 
\cite{CahnH74,TaylorCH92,Davis01}, and also \cite{ani3d,dendritic}.

We now non-dimensionalize the problem. As a length scale we choose $R$,
which we set to be $100\mu$m for snow crystal growth. 
As a time scale we choose 
$$
\tilde{t} = \frac{R^2}{\mathcal{D}}\, \frac{c_{\rm solid}}{c_{\rm sat}}\,.
$$
In addition, we non-dimensionalize the concentration by introducing
\begin{equation} \label{eq:uc}
u = \frac{c - c_{\rm sat}}{c_{\rm sat}}\,.
\end{equation}
Then, in terms of the new independent variables 
$\vec{\hat{x}} = \vec{x}/ R$ and $\hat{t} =
t /\tilde{t}$, 
we obtain (on dropping the $\,\hat{ \;}\,$ 
notation for the new variables for ease of exposition)
the equations
\begin{alignat}{2}
\frac{c_{\rm sat}}{c_{\rm solid}}\, \partial_t u - \Delta\, u & = 0 
\phantom{11111} &&\mbox{in }\,\, \Omega_+(t), \nonumber\\
\frac{\partial u}{\partial \vec\nu}
 &= {\cal V} \phantom{11111}  &&\mbox{on }\,\, \Gamma(t), \label{eq:1}\\[2mm]
\frac{\rho\,{\cal V}}{\beta(\vec\nu)} & = \alpha\,\kappa_\gamma
+ u \qquad &&\mbox{on }\,\, \Gamma(t), \label{eq:2} 
\end{alignat}
where $\rho := (\mathcal{D}\,c_{\rm sat})/(R\,{c_{\rm solid}}\,v_{\rm
  kin})$ and $\alpha:=\delta/R$. 
Since $c_{\rm sat} \ll c_{\rm solid}$,
we simplify the first equation to 
\begin{equation}\label{eq:3}
\Delta\, u = 0 \quad \mbox{in }\,\, \Omega_+(t)\,.
\end{equation}
We will choose $\gamma$ and $\beta$ of order one, and hence it will be important to
specify the order of magnitude of the quantities $\rho$ and
$\alpha$ in (\ref{eq:2}).  
Taking the values of $c_{\rm sat}/c_{\rm solid}$ and
$v_{\rm kin}$ from the table in \cite[p.~866]{Libbrecht05} into account,
we observe that
$$
\frac{c_{\rm sat}}{c_{\rm solid}\,v_{\rm kin}} \approx 
0.71\times 10^{-8}\,{\rm{s}}\,(\mu{\rm m})^{-1} 
$$
independently of the temperature $T$. Moreover, for
the time scale $\tilde t$, which depends on $c_{\rm sat}$, and hence on $T$,
we obtain a range from 100\,s at $-1^\circ$~C to
1300\,s at $-30^\circ$~C.
These time scales seem to be realistic when comparing with the
experiments reported in \cite{Libbrecht05,GondaY82}. 

For the diffusion constant of water vapour in air we take
$\mathcal{D}=2 \times 10^7 \,(\mu {\rm m})^2\,{\rm{s}}^{-1}$, see 
\citet[p.\ 866]{Libbrecht05}, 
which is valid at a pressure of $1\,$atm. With the values of
$\delta$ and $R$ mentioned further above we obtain 
\begin{equation}\label{param}
\rho \approx 1.42\times 10^{-3}, \qquad  \alpha=\delta/R \approx 10^{-5}\,.
\end{equation}

If not otherwise stated, we will always choose these parameters 
in all the snow crystal growth computations described in Section~\ref{eq:super}.
For the boundary condition we set, on recalling
(\ref{eq:nondimb}) and (\ref{eq:uc}), 
\begin{equation} \label{eq:uD}
u = \uD := \frac{c_{\rm super}}{c_{\rm sat}} \,\,
\mbox{ on }\,\, \partial\Omega\,.
\end{equation}
With the help of the table in \cite[p.~866]{Libbrecht05} we compute several
exemplary values for the fraction in (\ref{eq:uD}) for different values of
the temperature $T$ and the supersaturation $\varrho_{\rm super}$; see
Table~\ref{tab:uhat}.
\begin{table}
\center
\begin{tabular}{|r|c|c|c|c|c|c|c|}
\hline 
 & \multicolumn{6}{c|}{supersaturation $\varrho_{\rm super}$ (g/m$^3$) } \\
\cline{2-7} 
$T$~~~~~ & 0.01 & 0.02 & 0.05 & 0.1 & 0.2 & 0.3 \\ \hline
$-1^\circ$ C & $0.002$ & $0.005$ & $0.011$ & $0.023$ & $0.048$ & $0.069$ \\
$-2^\circ$ C & $0.003$ & $0.005$ & $0.012$ & $0.025$ & $0.049$ & $0.074$ \\
$-5^\circ$ C & $0.003$ & $0.006$ & $0.016$ & $0.031$ & $0.063$ & $0.094$ \\
$-10^\circ$ C& $0.005$ & $0.010$ & $0.024$ & $0.048$ & $0.095$ & $0.143$ \\
$-15^\circ$ C& $0.007$ & $0.015$ & $0.037$ & $0.074$ & $0.147$ & $0.221$ \\
$-30^\circ$ C& $0.030$ & $0.060$ & $0.150$ & $0.300$ & $0.601$ & $0.901$ \\
\hline
\end{tabular} 
\caption{Values of $\uD=c_{\rm super}/c_{\rm sat}$ depending on $T$ and
 $\varrho_{\rm super}$.}
\label{tab:uhat}
\end{table}%
In effect, we appear to have reduced the two parameter variation of the diagram
in Figure~\ref{fig:libbrecht} to the single parameter $\uD$ in (\ref{eq:uD}).
However, in our numerical simulations of snow crystal growth we will vary both
$\uD$ and the kinetic coefficient $\beta$. Although in reality not much is
known about the possible shapes and dependencies of $\beta$, it is known that 
$\beta$ strongly depends on $T$. Thus varying $\beta$ in our numerical
computations may be interpreted as simulating different (yet unknown) 
temperature regimes.

\section{Numerical method and anisotropies}\label{sec3}

For the numerical results in this paper we employ the finite element
approximation introduced by the authors in \cite{dendritic,crystal}
in order to approximate solutions of (\ref{eq:1}), (\ref{eq:2}),
(\ref{eq:3}), (\ref{eq:uD}). 
In the method a uniform time step $\tau>0$ is employed and 
the evolution of the crystal surface is tracked 
with the help of parametric meshes $\Gamma^h$ that are
independent from the bulk meshes $\mathcal{T}^h$ on which the approximation 
$u^h$ of $u$ is computed. 
The scheme uses an adaptive bulk mesh that has a fine mesh size
$h_f$ around $\Gamma^h$ and a coarse mesh size $h_c$ 
further away from it. Here
$h_{f} = \frac{2\,H}{N_{f}}$ and $h_{c} =  \frac{2\,H}{N_{c}}$
are given by two integer numbers $N_f >  N_c$, where we assume from now on that
$\Omega = (-H,H)^d$. The initial parametric mesh $\Gamma^h(0)$ consists of
$K^0_\Gamma$ vertices, and this mesh is locally refined, where elements become
too large during the evolution. 

In order to successfully model the
evolution of anisotropic interface evolution laws the authors
introduced a stable discretization in \cite{ani3d, dendritic}. We now
discuss how $\gamma$ and $\beta$ have to be chosen in order to model
situations with a hexagonal anisotropy. In this paper, we choose
surface anisotropies of the form 
\begin{equation} \label{eq:g1}
\gamma(\vec{p}) = \sum_{\ell=1}^L
\gamma_{\ell}(\vec{p}), \quad
\gamma_\ell(\vec{p}):= [{\vec{p} \cdot G_{\ell}\,\vec{p}}]^\frac12\,, 
\end{equation}
where $G_{\ell} \in \bR^{d\times d}$, for $\ell=1\to L$, 
are symmetric and positive definite matrices, and 
$\vec p = (p_1,\ldots,p_d)^T \in \bR^d$ denotes a vector in $\bR^d$.
We remark that anisotropies of the form (\ref{eq:g1}) admit a formulation
of $\kappa_\gamma$ in (\ref{eq:2}) which can be discretized
in a simple and stable way, see \cite{ani3d,dendritic}. 
We will now demonstrate that these forms
of $\gamma$ also allow one to model a hexagonal surface energy in a
simple way. To this end, let
$l_\epsilon(\vec{p}) := 
\left[ \epsilon^2\,|\vec{p}|^2  + p_1^2\,(1-\epsilon^2) \right]^{\frac12}
= \left[p_1^2 + \epsilon^2\,\sum_{i=2}^d p_i^2 \right]^{\frac12}$
for $\epsilon>0$.

\begin{figure}
\def\myheight{0.1\textheight}
\center
\includegraphics[angle=0,height=\myheight]{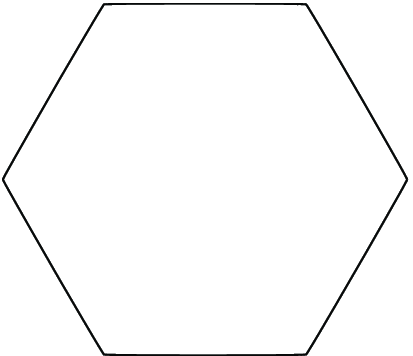} \qquad
\includegraphics[angle=0,height=\myheight]{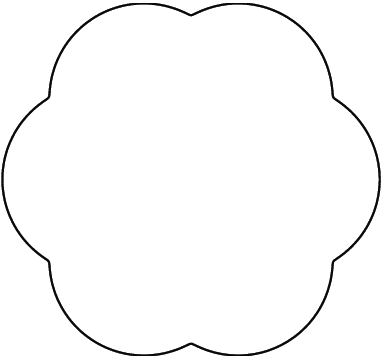}
\caption{Wulff shape (left) and polar plot (right) in $\bR^2$ 
for (\ref{eq:hexgamma2d}) with $\epsilon=0.01$ and $\theta_0=0$.}
\label{fig:Wulff2d} 
\end{figure}

Then a hexagonal anisotropy in $\bR^2$ can be modelled with the choice
\begin{equation} \label{eq:hexgamma2d}
\gamma(\vec{p}) =
\gamma_{hex}(\vec{p}) 
:= \sum_{\ell = 1}^3
l_\epsilon(R(\theta_0 + \frac{\ell\,\pi}3)\,\vec{p})\,,
\end{equation}
where $R(\theta)=
\left(\!\!\!\scriptsize
\begin{array}{rr} \cos\theta & \sin\theta \\
-\sin\theta & \cos\theta \end{array}\!\! \right)$ 
denotes a clockwise rotation through the angle $\theta$ and
$\theta_0 \in [0,\frac\pi3)$ is a parameter that rotates the orientation of
the anisotropy in the plane.
The Wulff shape 
of (\ref{eq:hexgamma2d}) for $\epsilon=0.01$ and $\theta_0=0$ 
is shown in Figure~\ref{fig:Wulff2d}, together with its polar plot 
$\mathcal{P} := \{ \gamma(\vec p)\,\vec p : |\vec p| = 1\}$. For more details 
on Wulff shapes and polar plots we refer to 
\cite{Davis01,Gurtin93}.

\begin{figure}
\def\mywidth{0.15\textwidth}
\center
\includegraphics[angle=0,width=\mywidth]{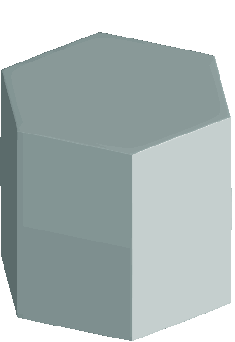} 
\qquad\qquad
\includegraphics[angle=0,width=\mywidth]{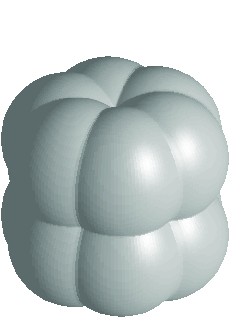} 
\caption{Wulff shape (left) and polar plot (right) 
in $\bR^3$ for (\ref{eq:L44}) with $\epsilon=0.01$.}
\label{fig:Wulff3d} 
\end{figure}

In order to define anisotropies of the form (\ref{eq:g1}) in $\bR^3$, we
introduce the rotation matrices
$R_{1}(\theta):=\left(\!\!\!\scriptsize
\begin{array}{rrr} \cos\theta & \sin\theta&0 \\
-\sin\theta & \cos\theta & 0 \\ 0 & 0 & 1 \end{array}\!\! \right)$ and 
$R_{2}(\theta):=\left(\!\!\!\scriptsize
\begin{array}{rrr} \cos\theta & 0 & \sin\theta \\
0 & 1 & 0 \\ -\sin\theta & 0 & \cos\theta \end{array}\!\! \right)$.
In this paper, we consider
\begin{equation} 
\gamma(\vec{p}) =
\gamma_{hex}(\vec{p})
:= l_\epsilon(R_2(\frac{\pi}2)\,\vec{p}) + \frac{1}{\sqrt{3}}
\sum_{\ell = 1}^3
l_\epsilon(R_1(\theta_0 + \frac{\ell\,\pi}3)\,\vec{p})\,,\label{eq:L44}
\end{equation}
which is relevant for the simulation of snow crystal growth.
Its Wulff shape for $\epsilon=0.01$ is shown in
Figure~\ref{fig:Wulff3d}, together with its polar plot.

We note that the Wulff shape of (\ref{eq:L44}) 
for $\epsilon\to0$ approaches a prism, where every face
has the same distance from the origin. In other words, for
(\ref{eq:L44}) 
the surface energy densities in the basal and prismal directions are the same.
We remark that if $\mathcal{W}_0$ denotes the Wulff shape of
(\ref{eq:L44}) with $\epsilon=0$, 
then the authors in \cite{GravnerG09} used the scaled
Wulff shape $\frac12\,\mathcal{W}_0$ as the building block in their 
cellular automata algorithm. In addition, we observe that the choice
(\ref{eq:L44}) agrees well with data reported in e.g.\ 
\cite[p.~148]{PruppacherK97}, although there the ratio of basal to prismal 
energy is computed as $\gamma^{\rm B} / \gamma^{\rm P} \approx 0.92 < 1$.
In order to be able to model this situation as well, we generalise the choice
(\ref{eq:L44}) to
\begin{equation}
\gamma(\vec{p}) = 
\gamma^{\rm TB}_{hex}(\vec{p}) 
:=
\gamma_{\rm TB}\,l_\epsilon(R_2(\frac{\pi}2)\,\vec{p})
 + \frac{1}{\sqrt{3}}\sum_{\ell = 1}^3
l_\epsilon(R_1(\theta_0 + \frac{\ell\,\pi}3)\,\vec{p})\,, \label{eq:L44gTB}
\end{equation}
so that now $\gamma^{\rm B} / \gamma^{\rm P} = \gamma_{\rm TB}$.

\begin{figure}
\def\myheight{0.1\textheight}
\center
\includegraphics[angle=0,height=\myheight]{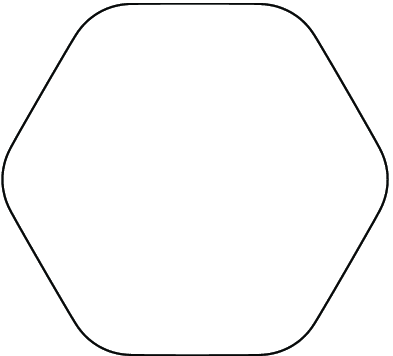} \qquad
\includegraphics[angle=0,height=\myheight]{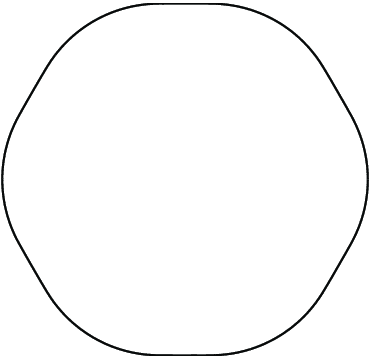}
\caption{Wulff shape in $\bR^2$ for (\ref{eq:hexgamma2ds}) with
$\sigma=1$ (left) and $\sigma=5$ (right).}
\label{fig:Wulff2ds} 
\end{figure}

A more generalized form of (\ref{eq:hexgamma2d}) and (\ref{eq:L44}), which also
fits into the framework (\ref{eq:g1}), is given by
\begin{equation} \label{eq:hexgamma2ds}
\gamma(\vec{p}) = \gamma_{hex} (\vec{p}) + \sigma\,|\vec p|\,,
\end{equation}
where $\sigma \geq 0$ is a fixed parameter. For the case $d=2$ we show the 
Wulff shape of (\ref{eq:hexgamma2ds}) for $\sigma=1$ and $\sigma=5$ in
Figure~\ref{fig:Wulff2ds}.
We note that for (\ref{eq:hexgamma2ds}) with $\epsilon=0$ and
$\sigma > 0$, both in the case $d=2$ and in the case $d=3$, the
corresponding Wulff shape still has flat parts, but is now smooth with no
corners and, if $d=3$, with no edges. 
Equilibrium crystal shapes with
these characteristics can be found in certain metals \cite{Bonzel03},
and it is conjectured that they may be relevant for snow crystals as well
\cite{Libbrecht_private}. 

As discussed in \cite{Libbrecht05}, and the references therein, the
precise values of $\beta$ as a function of the normal $\vec{\nu}$ are not
known. Hence one issue in our computations is to understand how
different choices of $\beta$ influence the overall evolution. First
choices for the anisotropy in the kinetic coefficient are
$\beta(\vec{\nu})\equiv 1$ and $\beta=\gamma$. It was discussed in
\cite{Libbrecht05} that the value of $\beta$ is expected to change
with temperature and can vary quite drastically as a function of the
orientation. Denoting by $\beta^B$ the condensation coefficient of the
basal directions and by $\beta^P$ the condensation coefficient in
the prismal directions, it is for example expected that the growth of thin
plates at $T=-15^\circ$C is only possible if $\beta^P/\beta^B$ is
large. 

In order to be able to vary the kinetic coefficient $\beta$
significantly, in the case $d=3$ we define for later use
\begin{equation} \label{eq:betaflat}
\beta_{\rm flat}(\vec{p}) = \beta_{\rm flat,\ell}(\vec{p}) 
:= [p_1^2 + p_2^2 + 10^{-2\ell}\,p_3^2]^\frac12\,,
\end{equation}
and
\begin{equation} \label{eq:betatall}
\beta_{\rm tall}(\vec{p}) = \beta_{\rm tall,\ell}(\vec{p}) 
:= [10^{-2\ell}\,(p_1^2 + p_2^2) + p_3^2]^\frac12
\end{equation}
with $\ell\in\mathbb{N}$. We note that in practice there is hardly any 
difference between the numerical results for a kinetic coefficient $\beta$ that
is isotropic in the $x_1-x_2$-plane, such as $\beta_{\rm flat}$ and 
$\beta_{\rm tall}$, and one that is anisotropically aligned to the surface
energy density, such as e.g.\ $\beta = \beta_{\rm flat}\,\gamma$. Hence in all
our three dimensional numerical simulations
we always choose coefficients $\beta$ that
are isotropic in the $x_1-x_2$-plane, e.g.\ (\ref{eq:betaflat}) or
(\ref{eq:betatall}). 

In addition, it might be the case that the condensation coefficient $\beta$
is considerably lower in the directions normal to the facets.
In order to model this we choose 
\begin{align}
\beta_{hex,L}(\vec{p}) &= 
\left(\beta_{\max}\,[\gamma_{hex}(\vec{p}) - \gamma_{\min}] + 
\beta_{\min}\,[\gamma_{\max} - \gamma_{hex}(\vec{p})]\right) 
\nonumber\\ & \qquad
/ (\gamma_{\max} - \gamma_{\min}) \,,
\label{eq:betaL}
\end{align}
where we fix $\beta_{\max} = 10^3$ and $\beta_{\min} = 1$, and where
$$
\gamma_{\max} := \max_{|\vec p|=1}\gamma_{hex}(\vec p) \in \Rgeq
\,\,\mbox{ and }\,\,
\gamma_{\min} := \min_{|\vec p|=1}\gamma_{hex}(\vec p) \in \Rgeq \,.
$$
We note that for the 2d anisotropy (\ref{eq:hexgamma2d}) it holds that
$\gamma_{\max} = \gamma_{hex}(e^{-{\rm i}\,\theta_0})$ and
$\gamma_{\min} = \gamma_{hex}(e^{{\rm i}\,(\frac\pi6 - \theta_0)})$.
For more details on the numerical method and the anisotropies 
we refer to \cite{dendritic,triplejANI,ani3d,crystal}.

\section{Numerical computations}\label{sec4}

\subsection{Snow crystal growth in two dimensions}

In all computations for (\ref{eq:1}), (\ref{eq:2}), (\ref{eq:3}),
(\ref{eq:uD}) in this subsection we will, if not otherwise stated, 
use the parameters
(\ref{param}) and choose the surface energy anisotropy $\gamma = \gamma_{hex}$
defined by (\ref{eq:hexgamma2d}) with $\epsilon=0.01$ and
$\theta_0=\frac{\pi}{12}$. The rotation in the definition of the
anisotropy is used, so that the dominant growth directions are not
exactly aligned with the underlying bulk meshes $\mathcal{T}^h$. 
Moreover, the radius of the circular initial crystal seed, $\Gamma(0)$, is
always chosen to be $0.05$. 

\begin{figure}
\center
\mbox{
\includegraphics[angle=-90,width=0.23\textwidth]{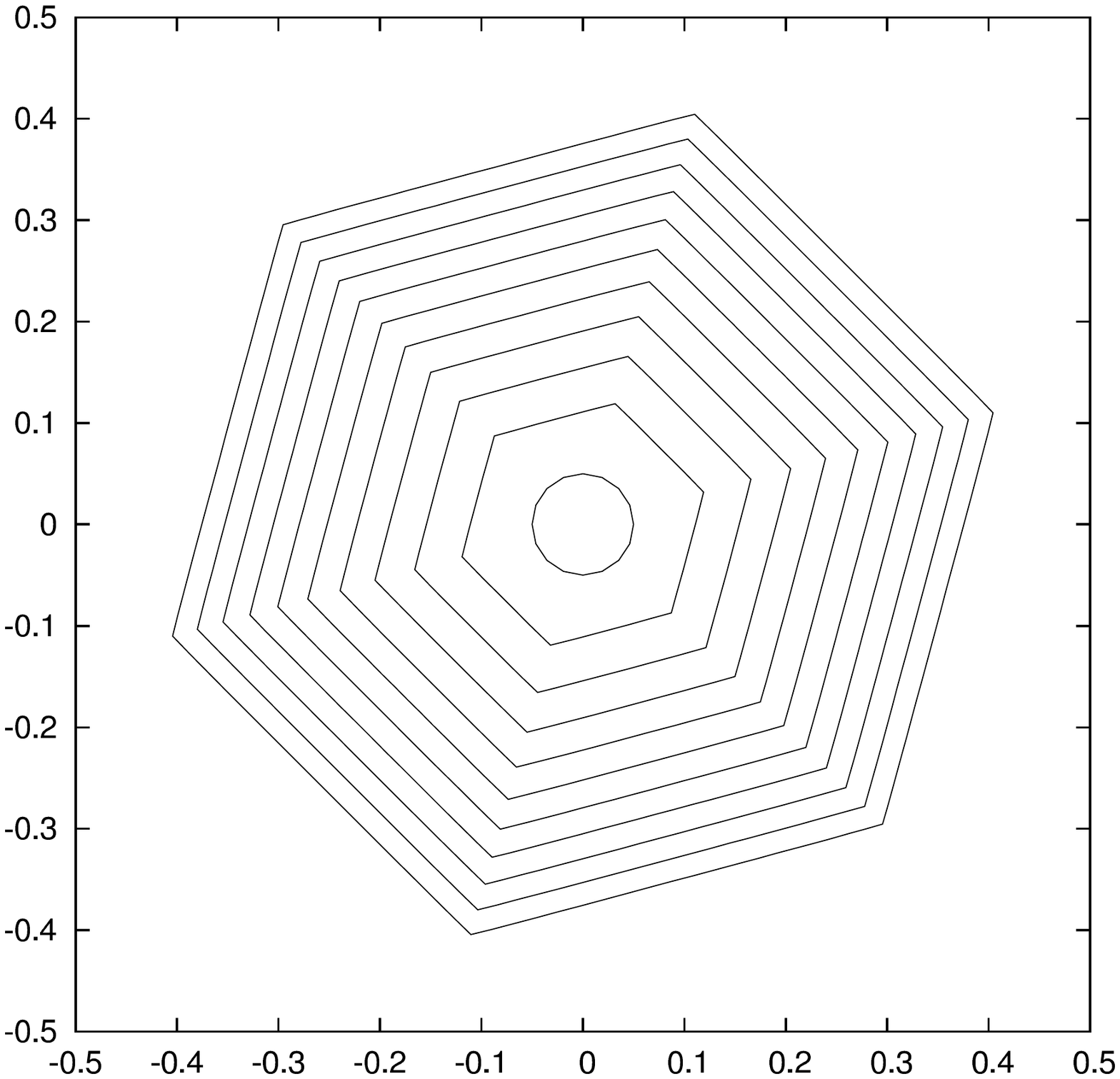}
\includegraphics[angle=-90,width=0.23\textwidth]{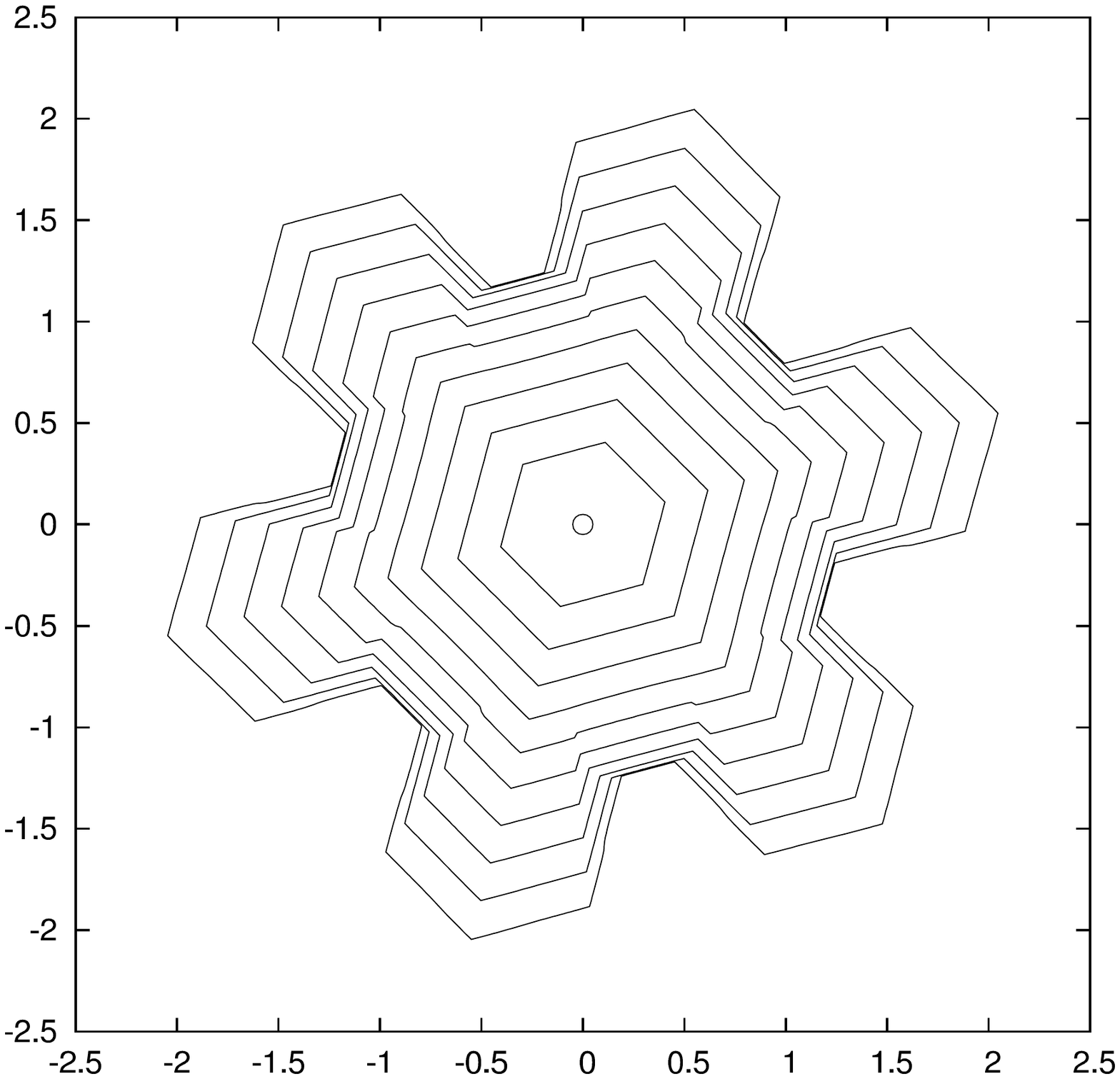}
}
\caption{($\Omega=(-4,4)^2$, $\uD = 0.004$, $\gamma=\beta = \gamma_{hex}$)
$\Gamma^h(t)$ for $t=0,\,5,\ldots,50$ (left) 
and for $t=0,\,50,\ldots,500$ (right).
Parameters are $N_f=256$, $N_c=4$, $K^0_\Gamma = 16$ and $\tau=0.1$.}
\label{fig:2dhex1} 
\end{figure}

First of all we study what influence the curvature and the velocity
terms in (\ref{eq:2}), and the supersaturation in (\ref{eq:uD}) 
have on the evolution of
the crystal. We choose the supersaturation $\uD=0.004$ and show the
results in Figure~\ref{fig:2dhex1}. One observes that the hexagonal
structure of the crystal forms quickly and that the facets become
unstable and break after they reached a certain size -- a phenomenon
which is observed in experiments as well, see
\cite{Libbrecht05}. 
\begin{figure}
\center
\mbox{
\includegraphics[angle=-90,width=0.23\textwidth]{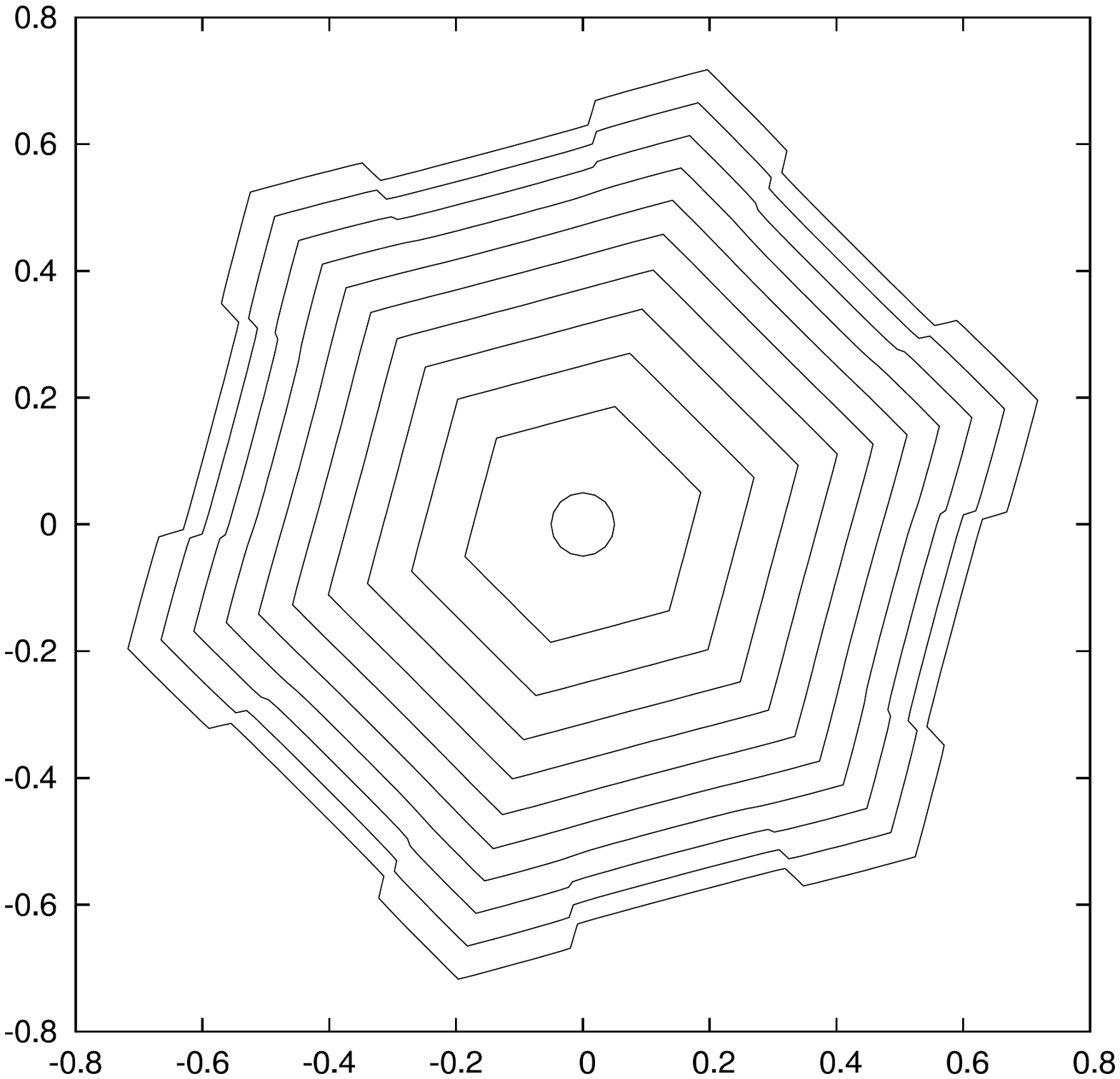}
\includegraphics[angle=-90,width=0.23\textwidth]{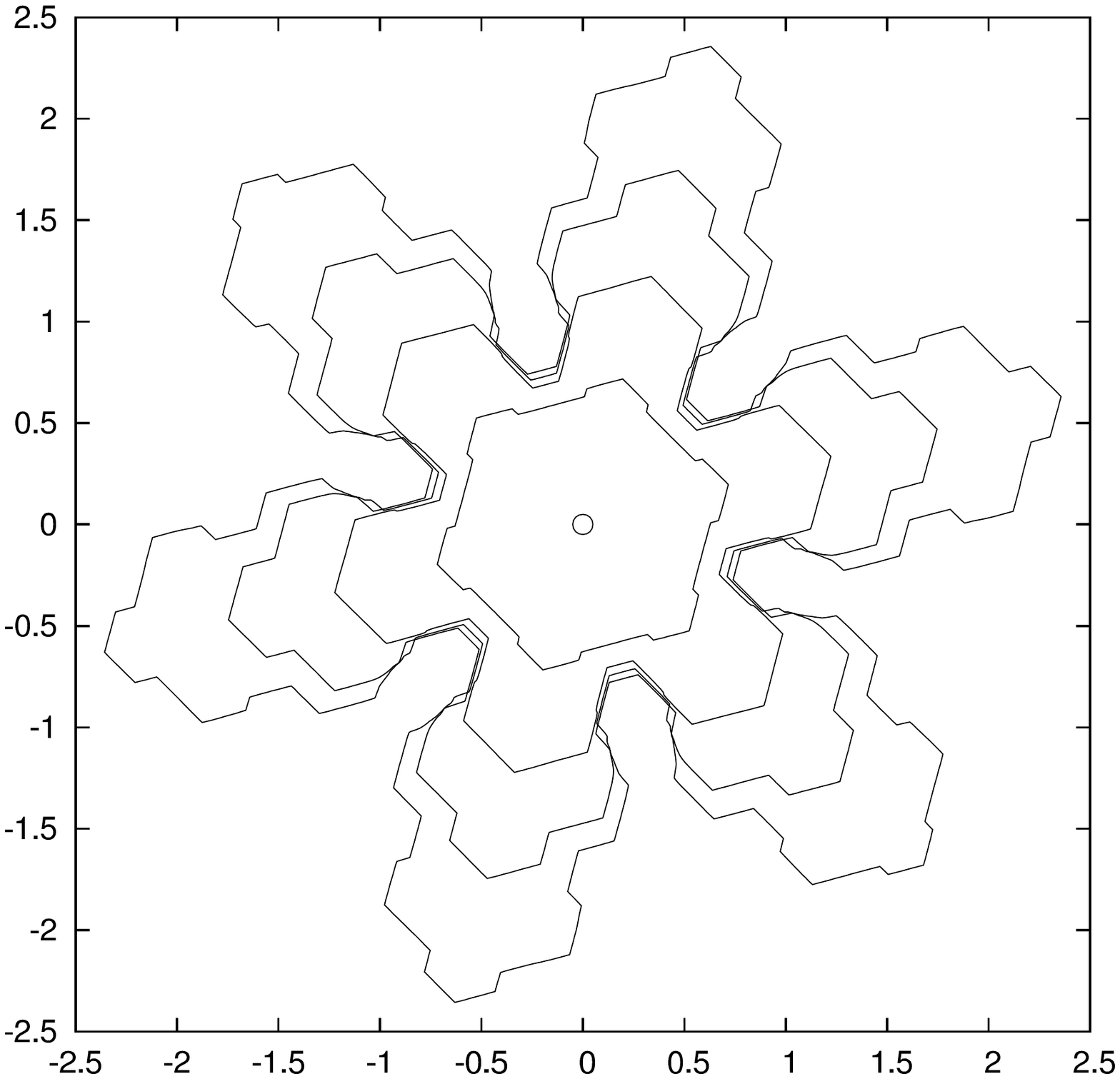}
}
\caption{($\Omega=(-4,4)^2$, $\uD = 0.01$, $\gamma=\beta=\gamma_{hex}$)
$\Gamma^h(t)$ for $t=0,\,5,\ldots,50$ (left), 
and for $t=0,\,50,\ldots,200$ (right).
Parameters are $N_f=512$, $N_c= K^0_\Gamma = 16$ and $\tau=5\times10^{-3}$.
}
\label{fig:2dhex2} 
\end{figure}%

\begin{figure}
\center
\mbox{
\includegraphics[angle=-90,width=0.23\textwidth]{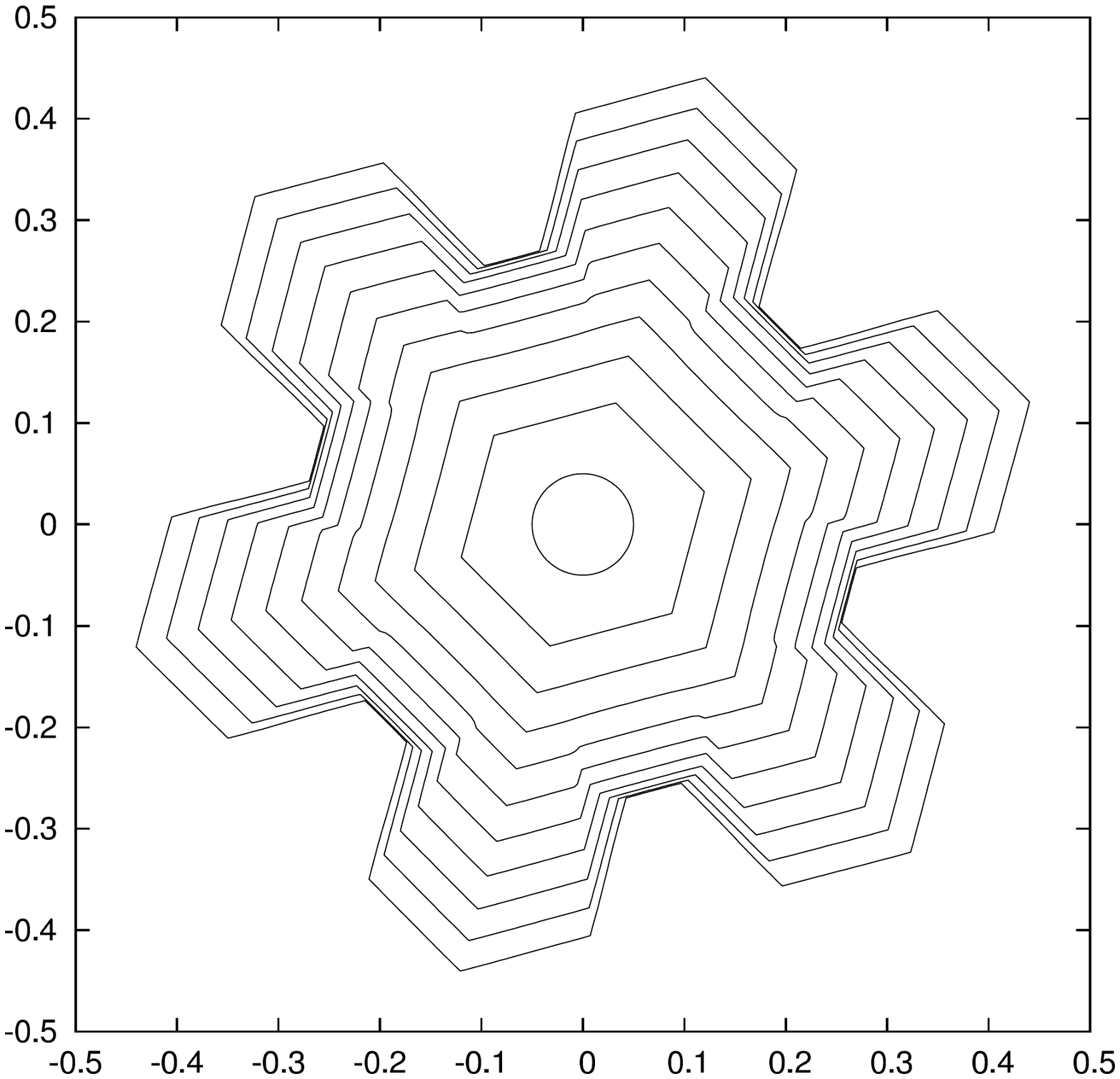}
\includegraphics[angle=-90,width=0.22\textwidth]{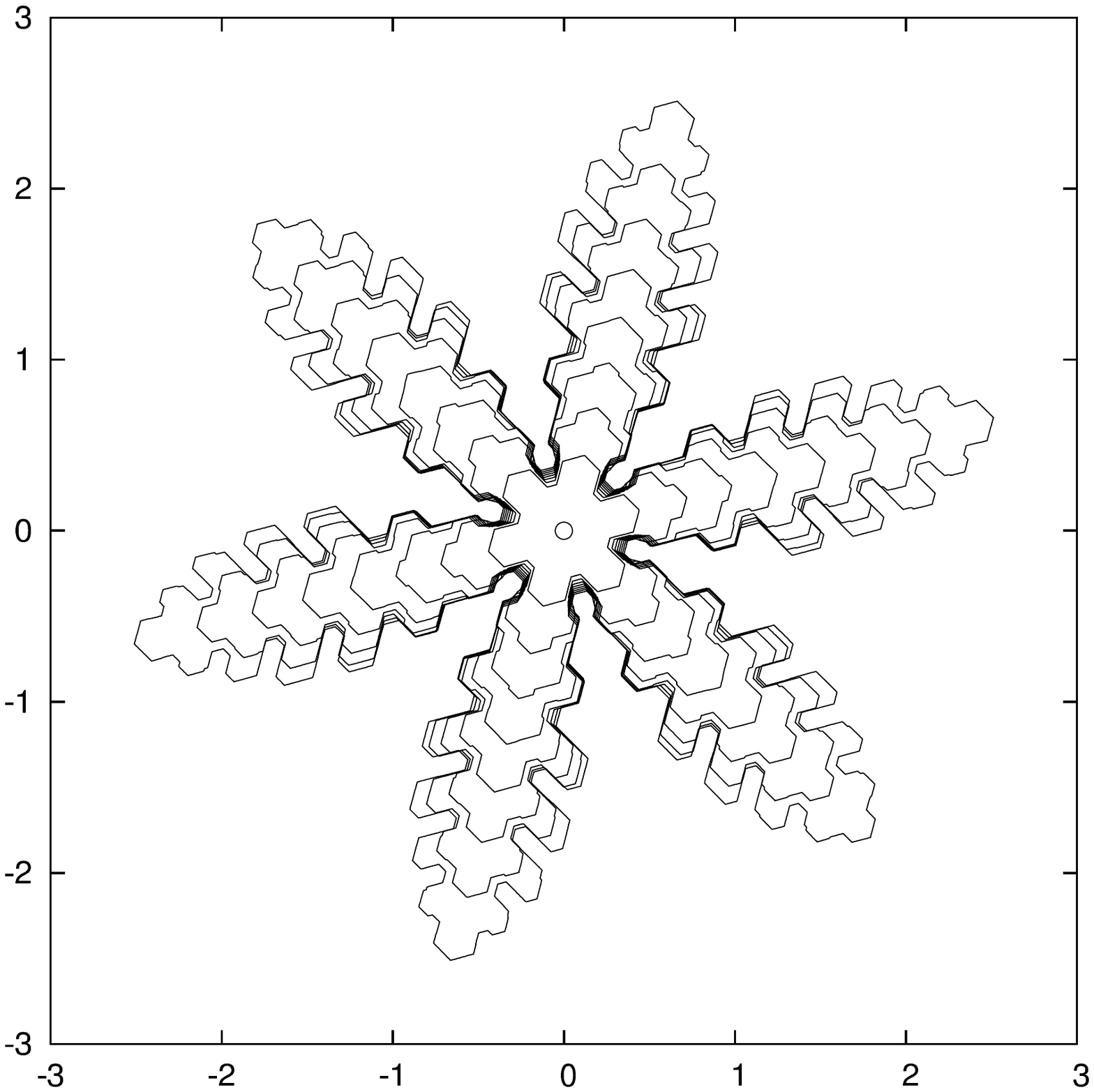}
}
\caption{($\Omega=(-4,4)^2$, $\uD   = 0.04$, $\gamma=\beta=\gamma_{hex}$)
$\Gamma^h(t)$ for $t=0,\,0.5,\ldots,5$ (left), 
and for $t=0,\,5,\ldots,40$ (right).
Parameters are $N_f=1024$, $N_c=K^0_\Gamma = 64$ and $\tau=2.5\times10^{-3}$.
}
\label{fig:2dhex3} 
\end{figure}%
\begin{figure}
\center
\mbox{
\includegraphics[angle=-90,width=0.23\textwidth]{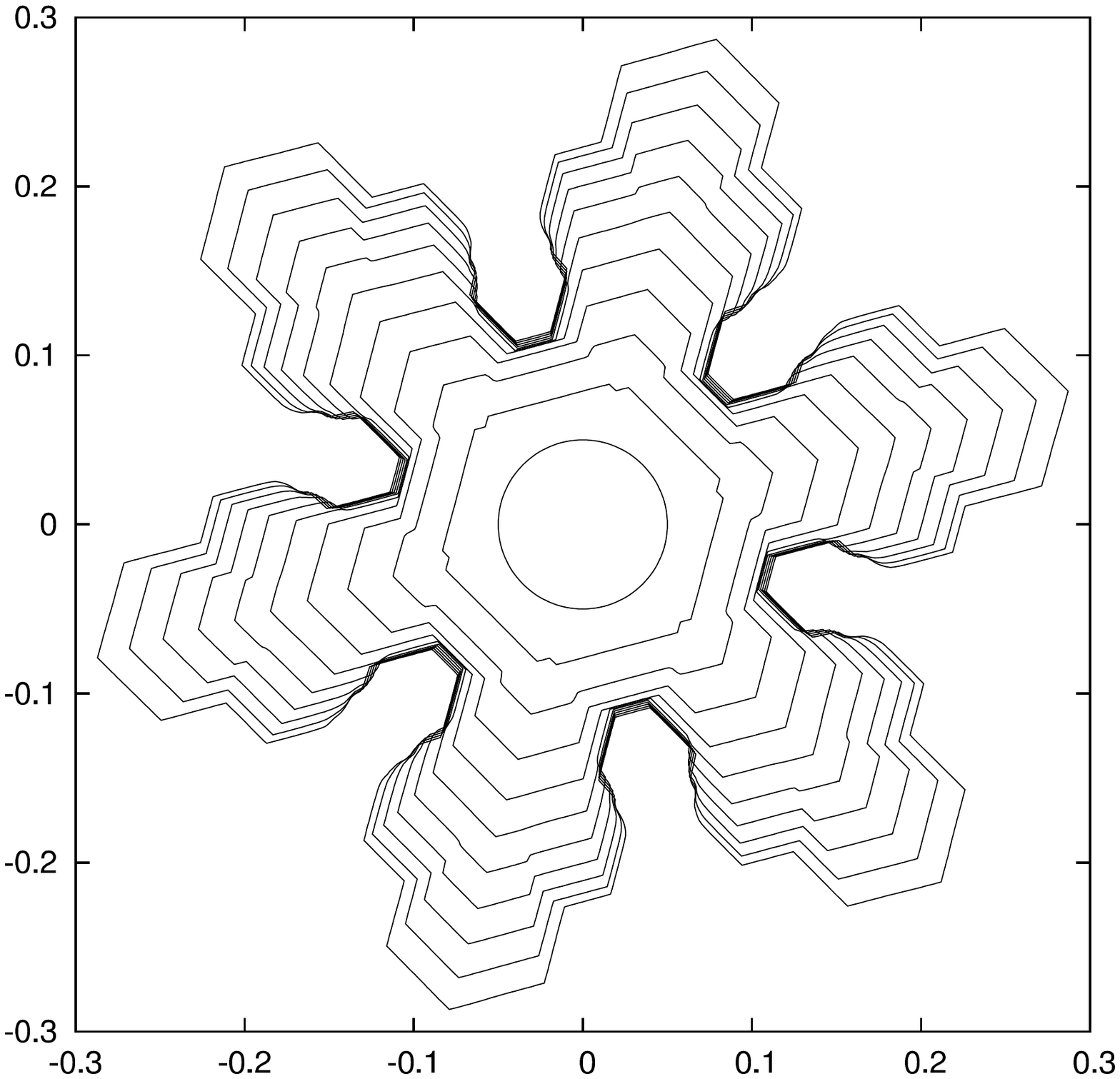}
\includegraphics[angle=-90,width=0.22\textwidth]{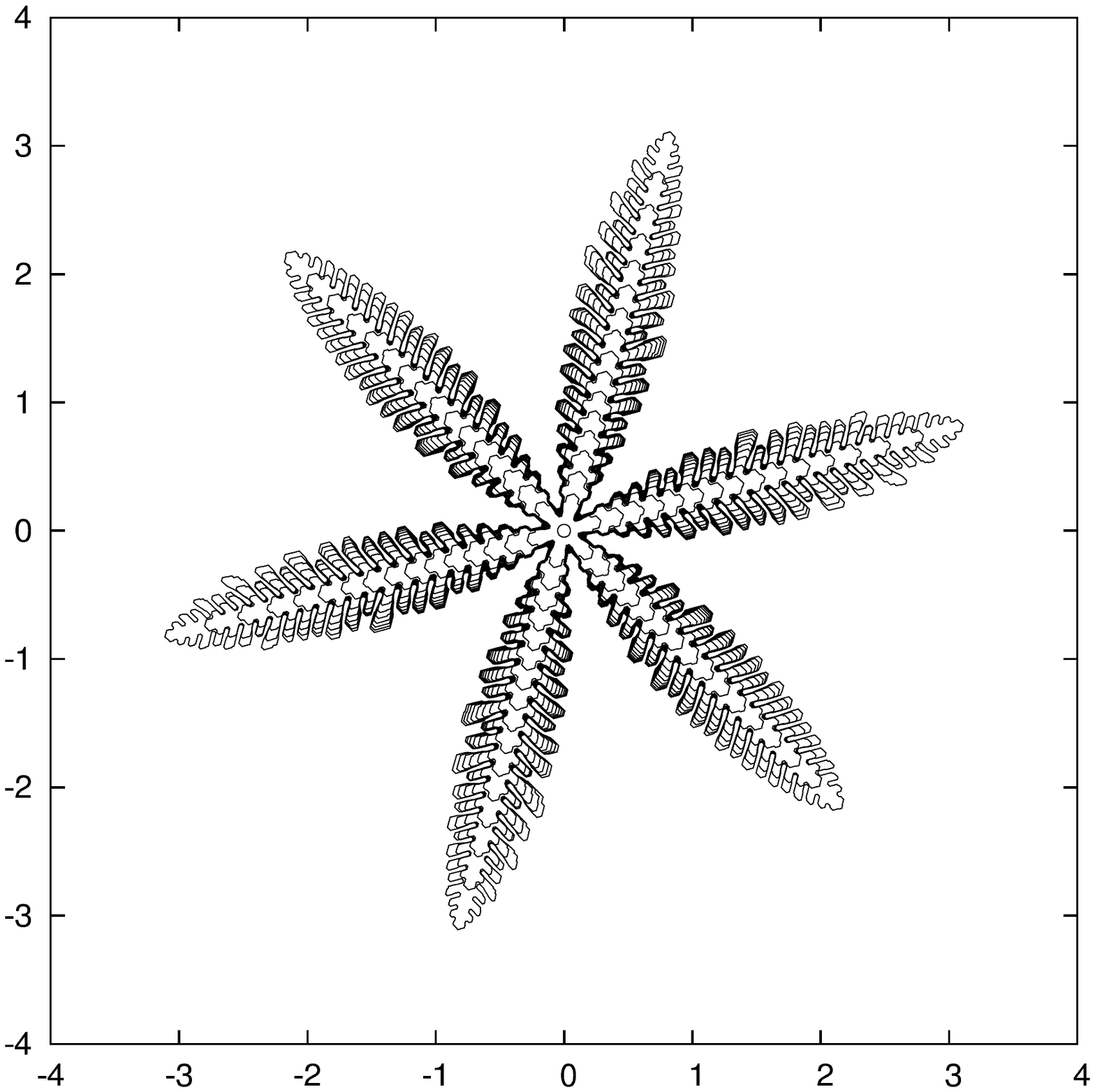}
}
\caption{($\Omega=(-4,4)^2$, $\uD = 0.2$, $\gamma=\beta=\gamma_{hex}$)
$\Gamma^h(t)$ for $t=0,\,0.04,\ldots,0.4$ (left), 
and for $t=0,\,0.4,\ldots,6.4$ (right).
Parameters are $N_f=2048$, $N_c = K^0_\Gamma = 128$ 
and $\tau=2.5\times10^{-4}$.}
\label{fig:2dhex4} 
\end{figure}%

In Figures~\ref{fig:2dhex2}--\ref{fig:2dhex4} we plot computations with larger
supersaturations, $\uD=0.01, 0.04, 0.2$. One clearly observes
dendritic growth, which is more enhanced at larger supersaturations. In
addition, the evolution is much faster due to the fact that more water
vapour molecules are available. 

\begin{figure}
\center
\includegraphics[angle=-90,width=0.4\textwidth]{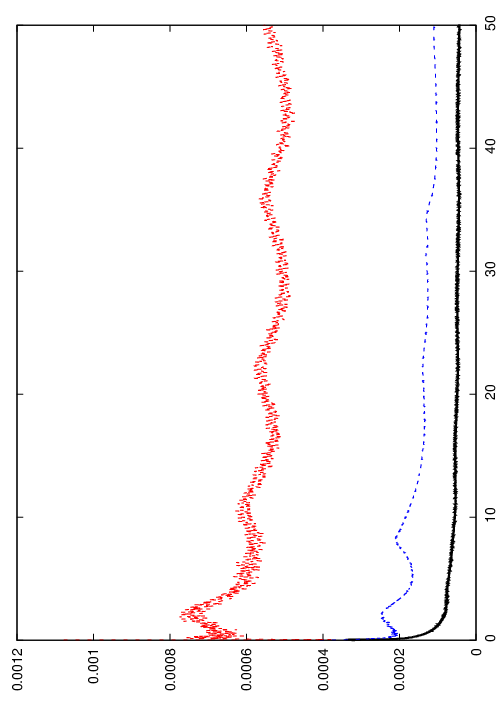}
\caption{($\Omega=(-8,8)^2$, $\uD = 0.04$, $\gamma=\gamma_{hex}$, $\beta = 1$)
Approximations of 
$\rho\,\widehat{\mathcal{V}}$ (black, solid), 
$\alpha\,\kappa_\gamma^{\rm avg}$ (blue, dashed) and 
$\alpha\,\kappa_\gamma^{\max}$ (red, dashed). Here $\rho$ and $\alpha$
are as in (\ref{param}). 
The time interval is $[0,50]$.
}
\label{fig:comparison1} 
\end{figure}

It is also of interest to compare the size of the terms appearing in 
(\ref{eq:2}). To this end, we compare the numerical approximations of the
terms
$$\rho\widehat{\mathcal{V}},\,\,
\alpha\,\kappa^{avg}_\gamma,\,\,\alpha\,\kappa^{max}_\gamma$$
where $\widehat{\mathcal{V}}$ denotes the observed tip velocity, i.e.\ the
velocity of the part of the interface furthest away from the origin,
$\kappa^{avg}_\gamma$ is the average of $|\kappa_\gamma|$ on the
interface and $\kappa^{max}_\gamma$ is the maximum of
$|\kappa_\gamma|$. For a computation with $\gamma=\gamma_{hex}$,
$\beta=1$ and $\uD=0.04$ we plot these values in
Figure~\ref{fig:comparison1}. 
It clearly can be seen that the curvature contribution is
larger than the velocity term. 
In our other computations we only observed
for supersaturations around $0.2$ and larger that the velocity term 
$\rho\widehat{\mathcal{V}}$ is larger than the average curvature term
$\alpha\,\kappa^{avg}_\gamma$. 

\begin{figure}
\center
\mbox{
\includegraphics[angle=-90,width=0.23\textwidth]{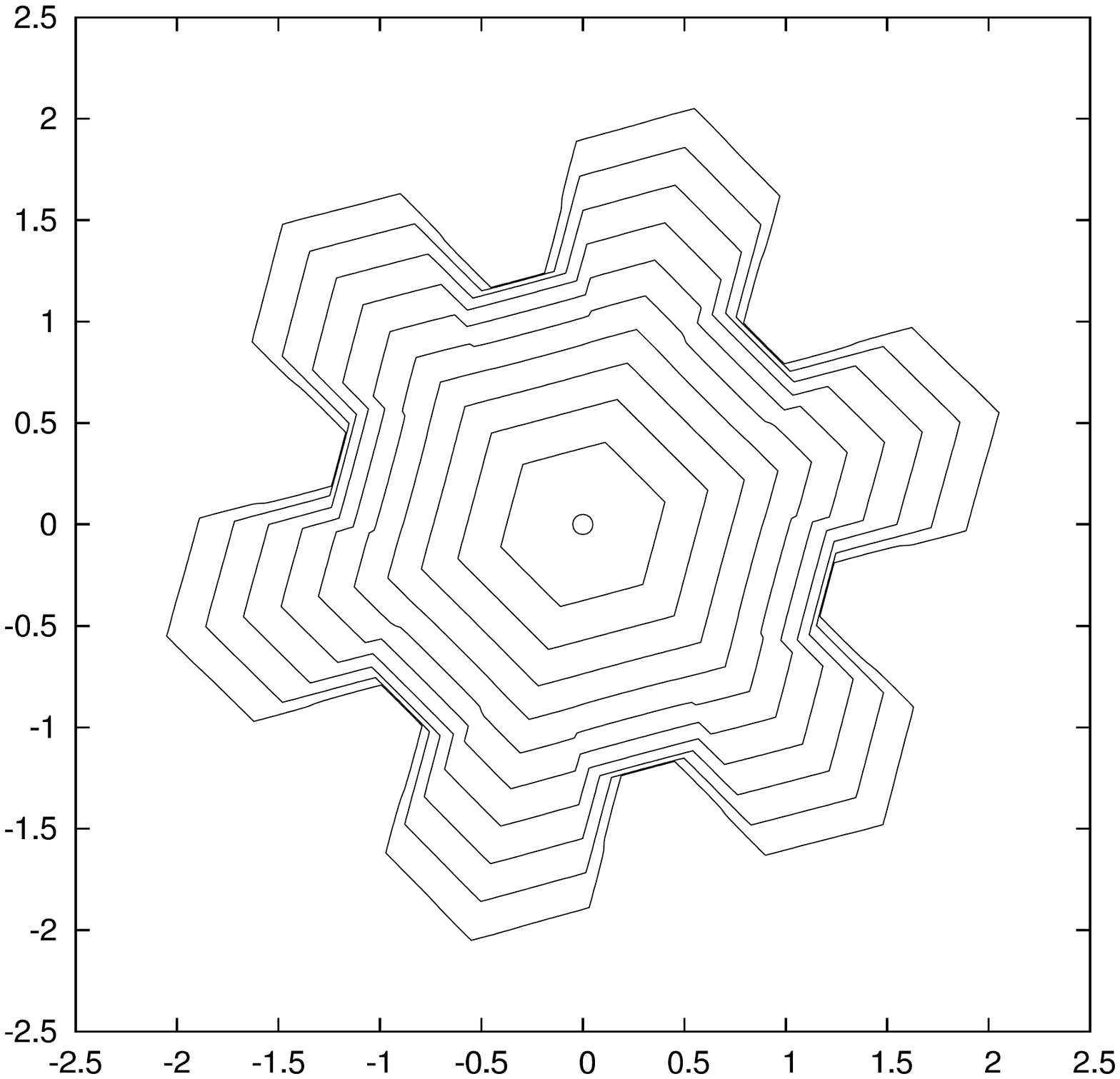}
\includegraphics[angle=-90,width=0.24\textwidth]{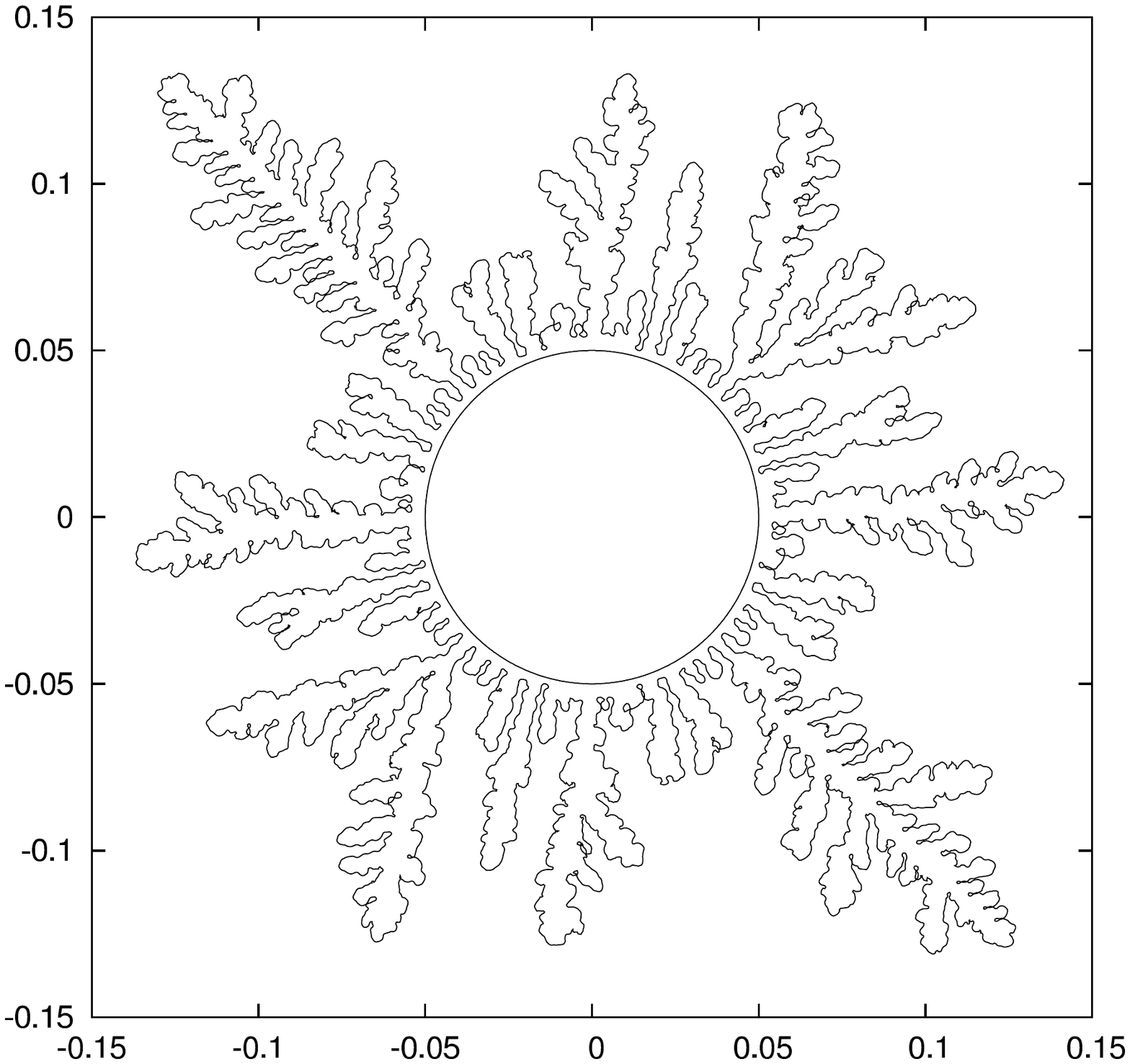}
}
\caption{($\Omega=(-4,4)^2$, $\uD = 0.004$) We choose
  $\gamma=\gamma_{hex}$, $\rho=0$ with parameters $N_f=256$, $N_c=4$,
  $K^0_\Gamma=16$ and $\tau=0.1$ and 
  plot $\Gamma^h(t)$ for $t=0,50,\ldots,500$ on the
  left, and we choose $\alpha = 0$, $\beta = \gamma_{hex}$ with parameters $N_f
  =  2048$, $N_c =128$, $K^0_\Gamma = 1024$ and $\tau=10^{-3}$ and plot 
  $\Gamma^h(t)$ for $t=0,\,3$ on the right.
}
\label{fig:comparison3} 
\end{figure}%

In Figure~\ref{fig:comparison3} we set the velocity term to zero
in the left computation, i.e.\ $\rho=0$, and we set the
curvature term to zero in the right computation, i.e.\ $\alpha=0$. 
We observe that leaving out the velocity term only
has a very minor impact on the crystal evolution. On the other hand,
leaving out the curvature term has a drastic effect -- the front becomes
very unstable. This can be explained as follows. 
Growth from supersaturated vapour is unstable and 
while the velocity term in (\ref{eq:2})
without the curvature term can dampen the unstable modes,
they are still unstable on all wavelengths. 
Whereas, the curvature term will stabilise the
small wavelengths and will select a fastest growing wavelength,
irrespective of the velocity term.

\begin{figure}
\center
\mbox{
\includegraphics[angle=-90,width=0.23\textwidth]{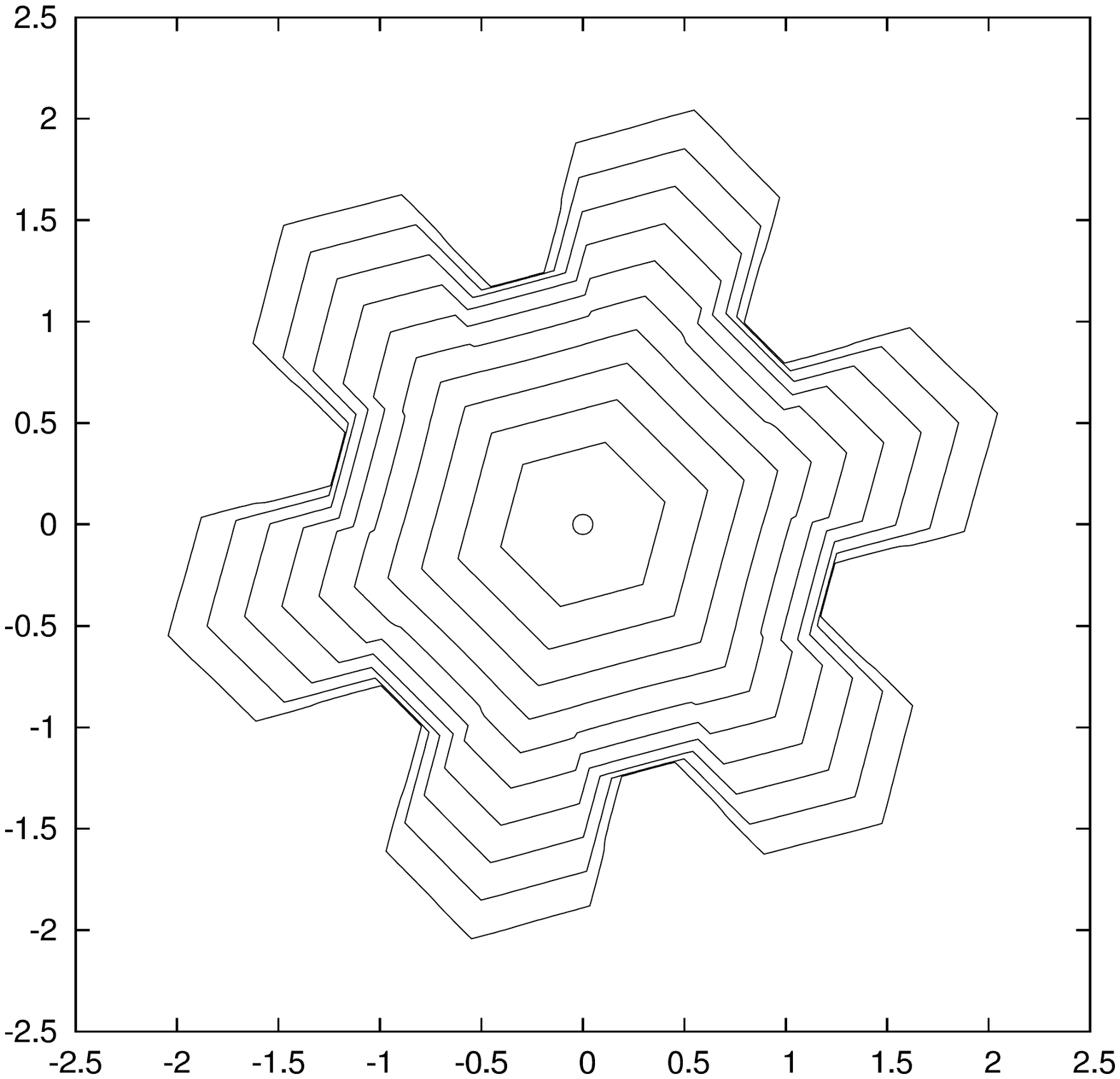}
\includegraphics[angle=-90,width=0.23\textwidth]{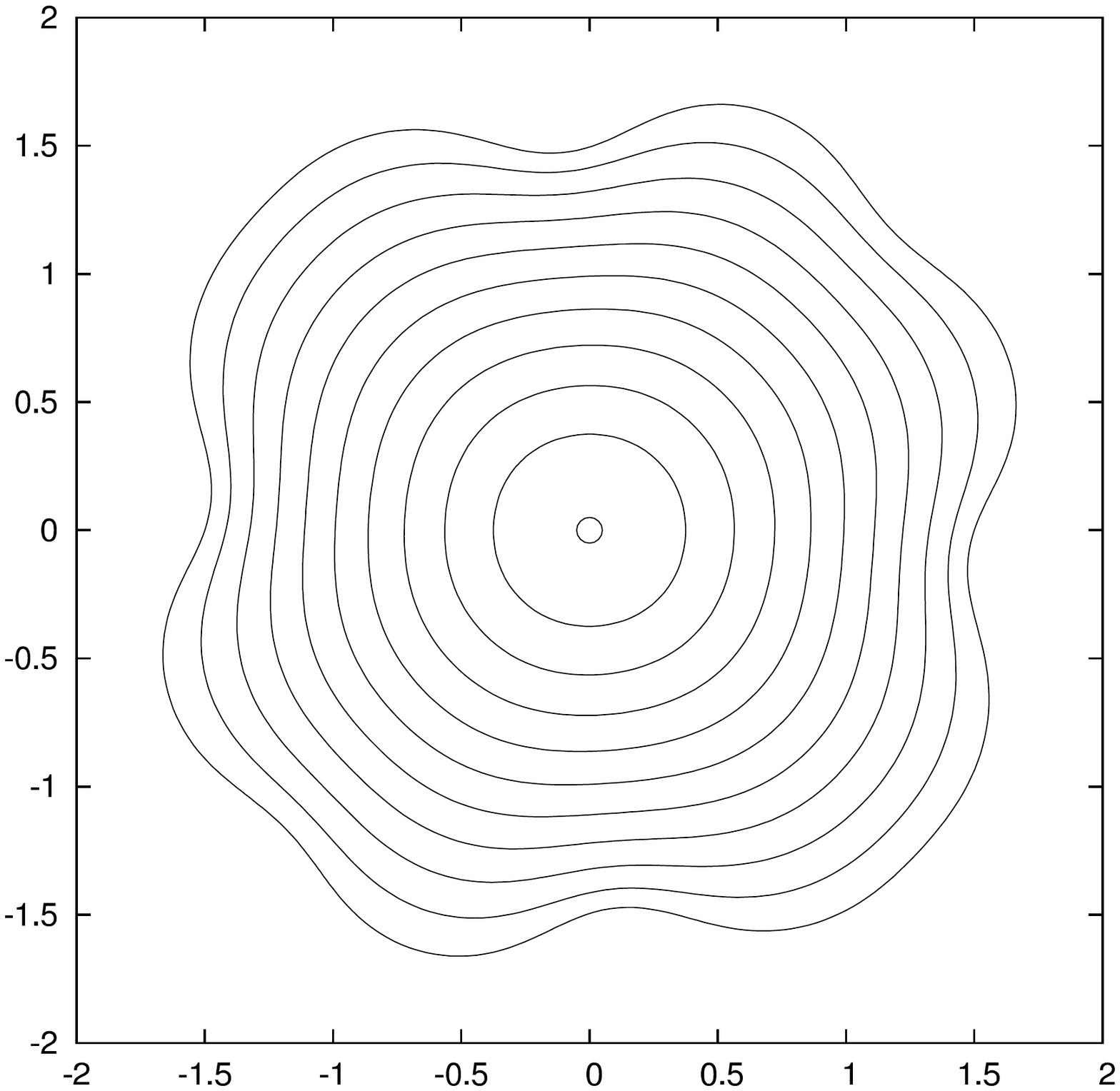}
}
\caption{($\Omega=(-4,4)^2$, $\uD = 0.004$) We take
  $\gamma=\gamma_{hex}$, $\beta = 1$, $N_f=256$, $N_c=4$, $K^0_\Gamma
  = 16$ and $\tau=0.1$ on the left and plot $\Gamma^h(t)$ for
  $t=0,\,50,\ldots,500$. We take $\gamma=\gamma_{iso}$,
  $\beta=\gamma_{hex}$, $N_f=512$, $N_c=16$, $K^0_\Gamma = 16$ and
  $\tau=10^{-2}$ on the right and plot $\Gamma^h(t)$ for $t=0,\,50,\ldots,500$. 
}
\label{fig:comparison2} 
\end{figure}

We now study the influence of the anisotropy on the evolution. 
On the left of Figure~\ref{fig:comparison2} we present an evolution with
a hexagonal anisotropy for $\gamma$ and an isotropic $\beta$. In the
same figure on the right we take $\gamma$ isotropic and $\beta$
hexagonal. One clearly observes
that anisotropy in the surface energy seems to be important to obtain
facetted growth. 

\begin{figure}
\center
\includegraphics[angle=-90,width=0.23\textwidth]{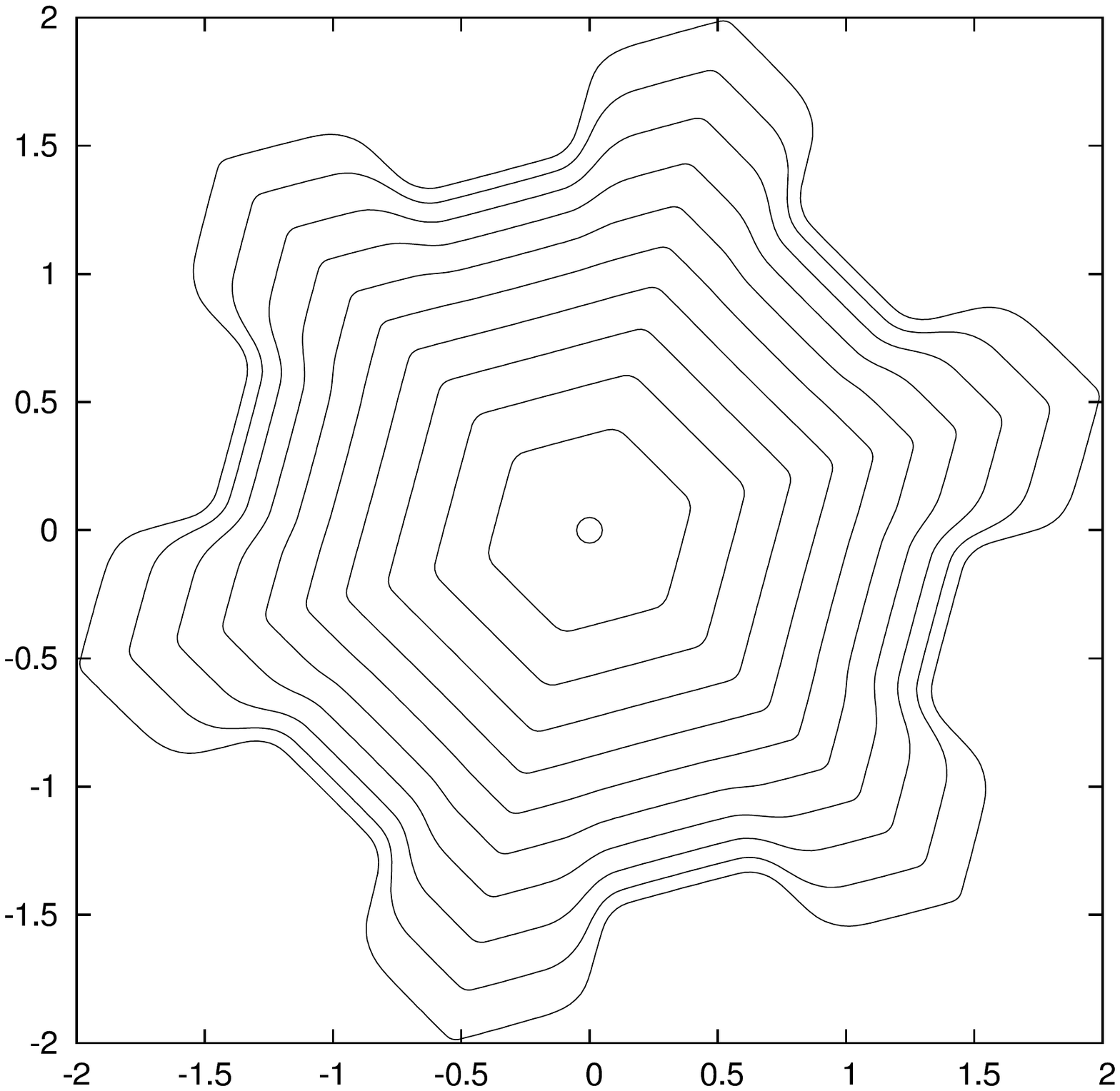}
\includegraphics[angle=-90,width=0.23\textwidth]{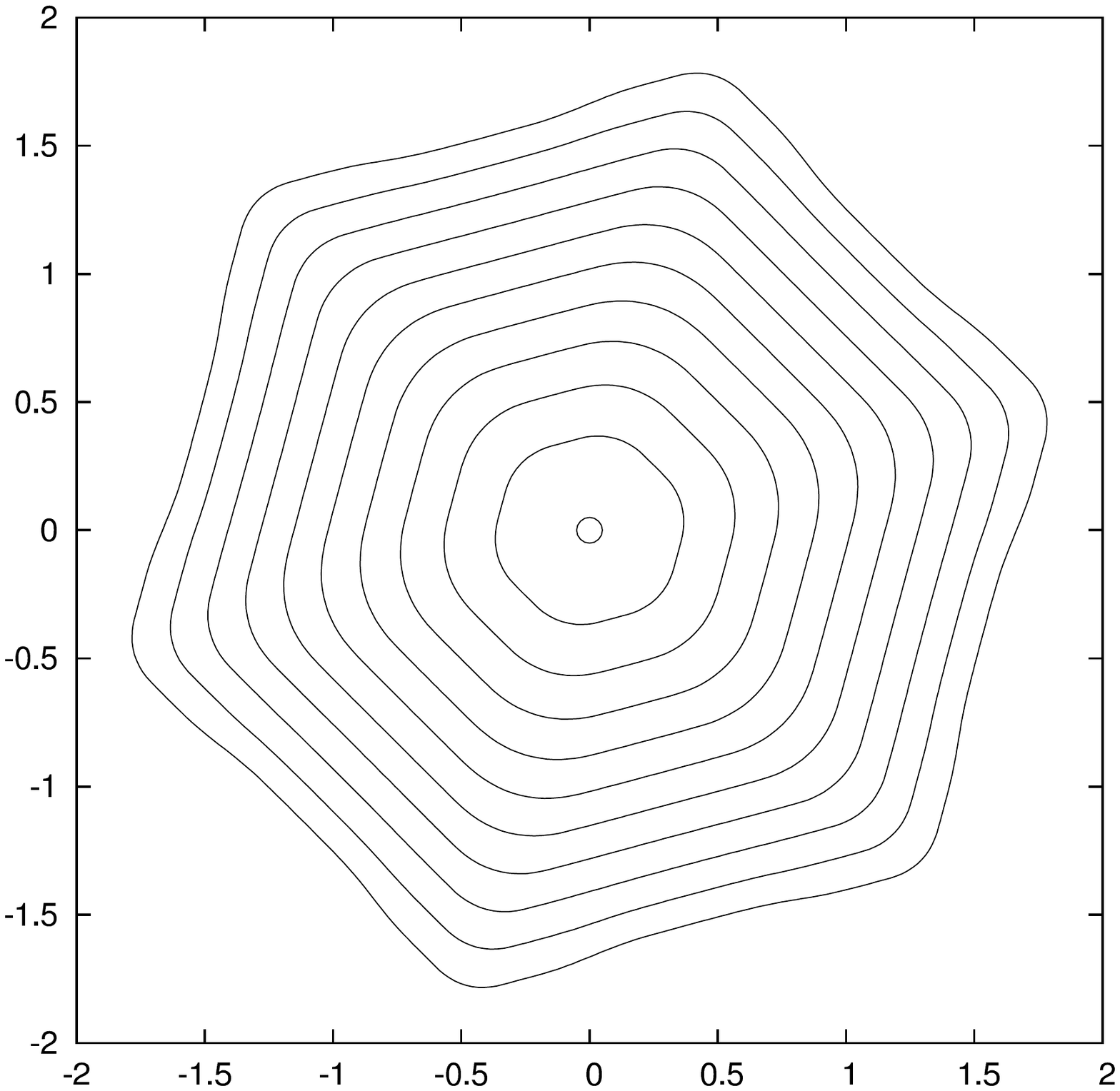}
\caption{($\Omega=(-4,4)^2$, $\uD = 0.004$, $\gamma$ as in 
(\ref{eq:hexgamma2ds}) with $\sigma=1$ (left) and $\sigma=5$ (right), 
$\beta = \gamma_{hex}$)
$\Gamma^h(t)$ for $t=0,\,50,\ldots,500$.
Parameters are $N_f=256$, $N_c=4$, $K^0_\Gamma = 16$ and $\tau=0.1$.
}
\label{fig:2dhex_L13} 
\end{figure}%
\begin{figure}
\center
\includegraphics[angle=-90,width=0.30\textwidth]{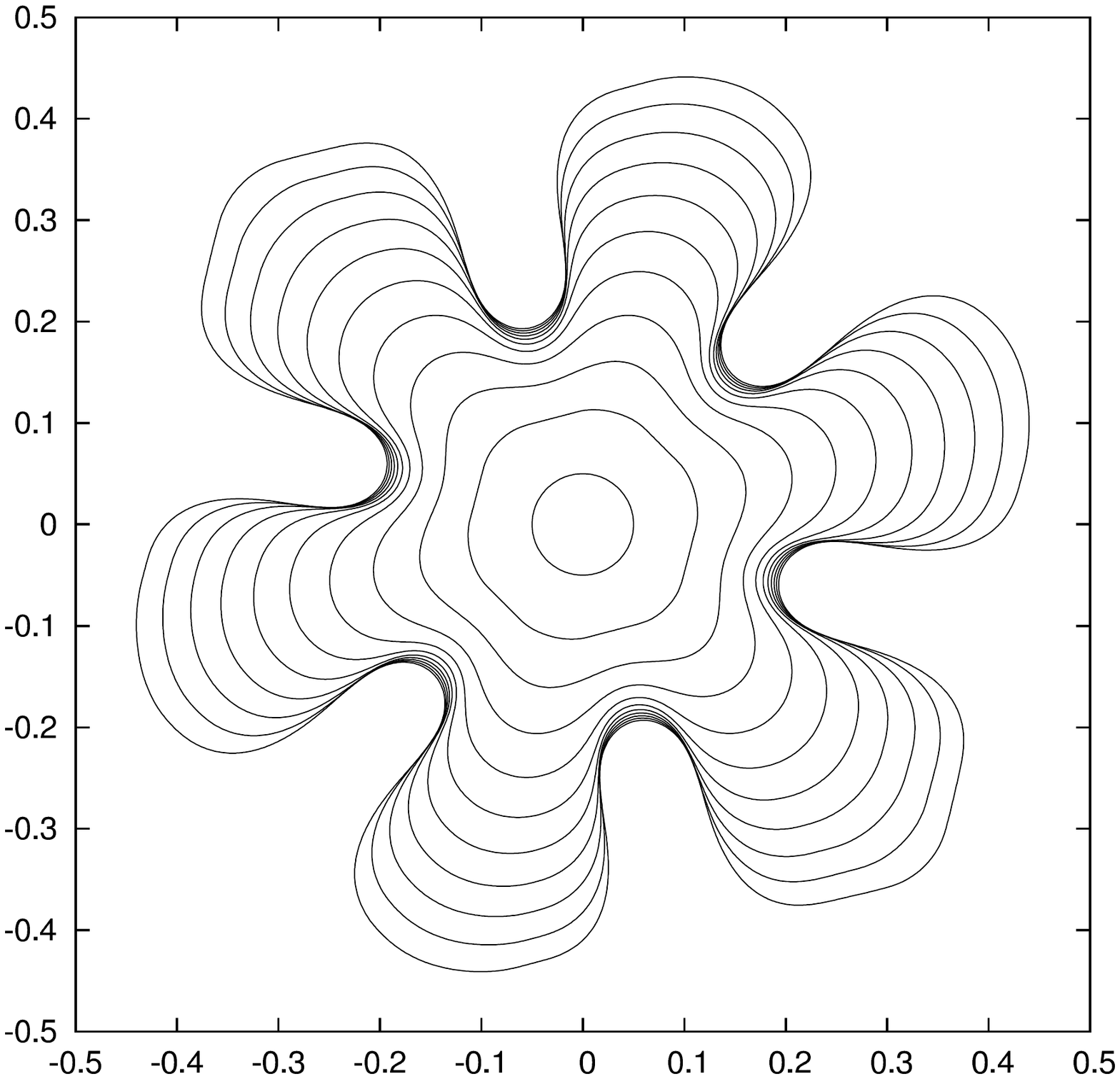}
\caption{($\Omega=(-4,4)^2$, $\uD = 0.04$, $\gamma=\gamma_{iso}$, 
$\beta=\beta_{hex,L}$) $\Gamma^h(t)$ for $t=0,\,0.5,\ldots,5$.
Parameters are $N_f=2048$, $N_c= K^0_\Gamma = 256$ and $\tau=10^{-3}$.
}
\label{fig:2dlargebetadiff} 
\end{figure}%
\begin{figure}
\center
\includegraphics[angle=-90,width=0.45\textwidth]{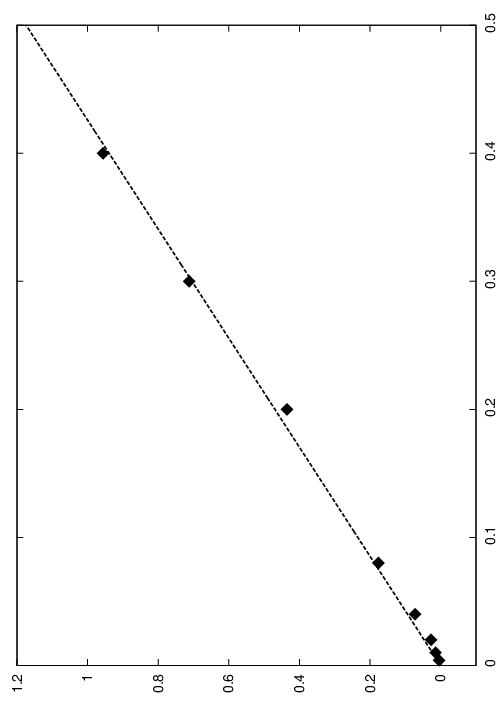}
\caption{($\Omega=(-4,4)^2$, $\gamma=\gamma_{hex}$, $\beta = 1$)
Best linear fit for the tip velocity $\widehat{\mathcal{V}}$ against the
supersaturation $\uD$.
}
\label{fig:2dvelofit} 
\end{figure}%
This is underlined by the next computation, where we choose 
(\ref{eq:hexgamma2ds}) for the anisotropy $\gamma$, and let 
$\beta = \gamma_{hex}$.
In Figure~\ref{fig:2dhex_L13} we present evolutions for
$\sigma=1$ and $\sigma=5$. These computations show 
how important the facetted anisotropy in the                           
Wulff shape is in order to obtain facetted snow crystals,
recall Figure~\ref{fig:Wulff2ds}.

It was suggested to the authors by Prof.\ Libbrecht that the difference of
$\beta$ as a function of $\vec{\nu}$ might be large with a minimum in
the the directions of the facet normals. 
We hence took $\beta=\beta_{hex,L}$ and
$\gamma=\gamma_{iso}$ in Figure~\ref{fig:2dlargebetadiff}. One
observes that the anisotropy in the condensation coefficient $\beta$ 
is large enough to lead to a six folded branching structure. But it is also
clearly visible that the anisotropy in $\beta$ does not lead to
facetted growth.

Next we study how the tip velocity in a growing dendritic crystal
depends on the supersaturation. So far nothing is known theoretically for
this dependence in the case of facetted growth, see 
\cite[\S4]{Libbrecht05}, even though for simpler problems, i.e.\ 
in the absence of facetting,
a vast literature exists, see e.g.\ \cite{Davis01} and the references therein. 
As can be seen in
Figure~\ref{fig:comparison1}, the tip velocity after some time becomes
basically time-independent. This is true also for our other
computations with different supersaturations. We observe a linear dependence between the
supersaturation and the tip velocity that the evolution eventually
settles on. To underline this qualitative behaviour, we numerically determine
the value for the nearly constant tip velocity $\widehat{\mathcal{V}}$ for 
several
values of $\uD$, see Figure~\ref{fig:2dvelofit} for a plot of these velocities.
We also show the best linear fit to this data, which is given by a
linear function with slope $\approx 2.35$. We remark that a similar
linear relationship between velocity and supersaturation has been observed
experimentally for needles, see \cite{LibbrechtCS02}. 

\begin{figure}
\center
\mbox{
\includegraphics[angle=-90,width=0.23\textwidth]{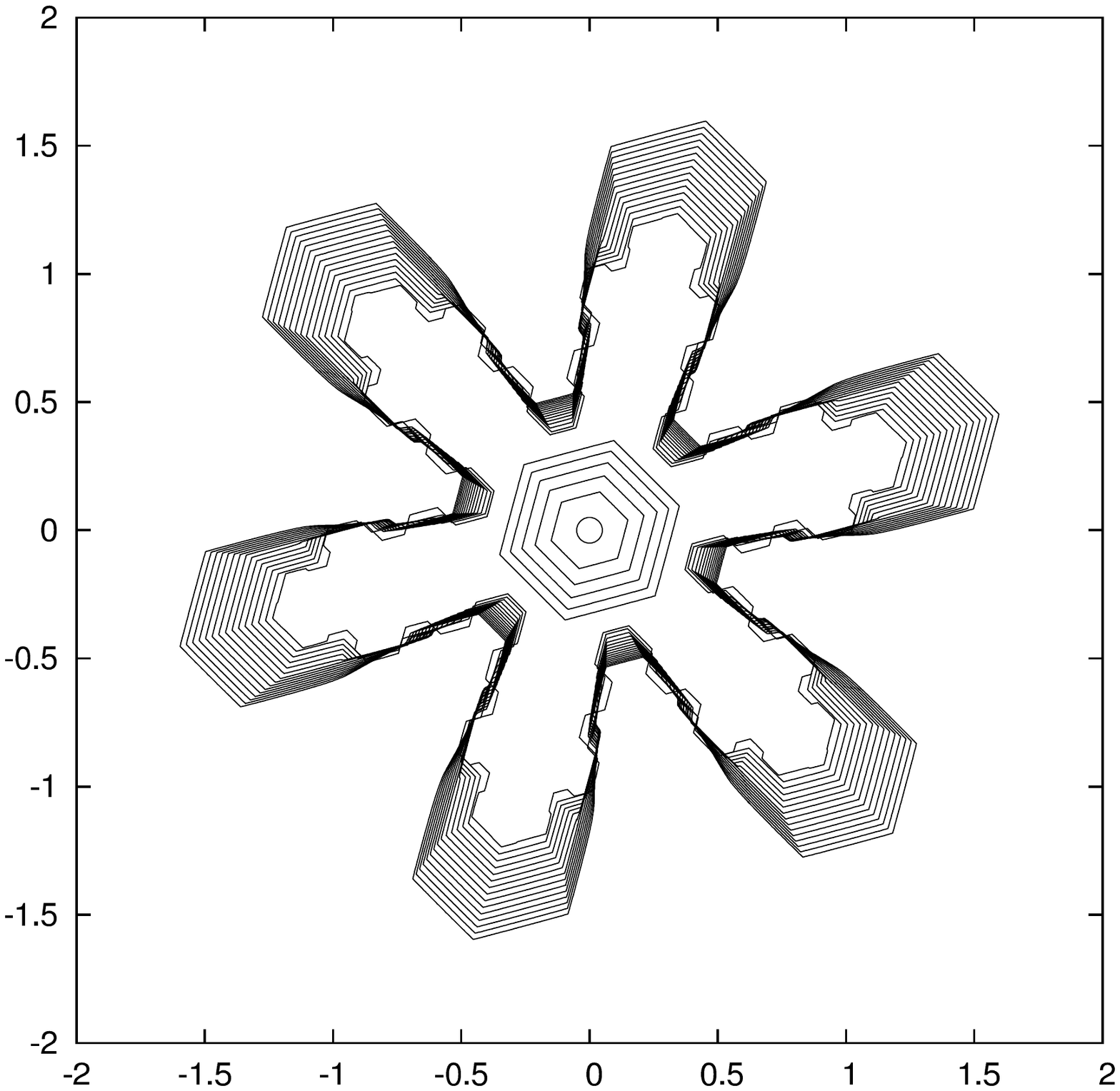}
\includegraphics[angle=-90,width=0.22\textwidth]{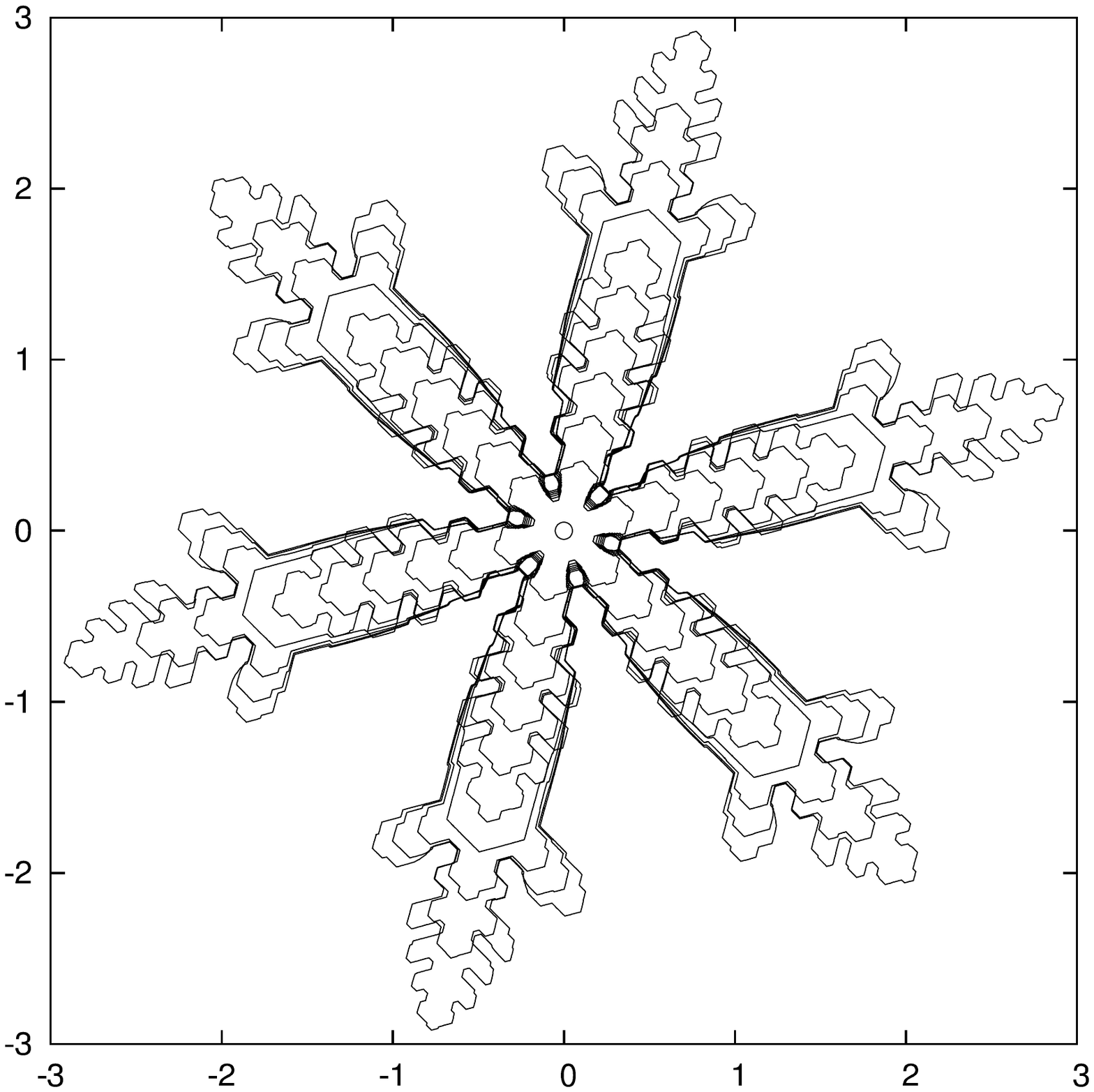}
}
\caption{($\Omega=(-4,4)^2$, $\gamma=\beta=\gamma_{hex}$)
$\Gamma^h(t)$ for $t=0,\,10,\ldots,200$ (left, (\ref{eq:uDboost})) and for
$t=0,\,2,\ldots,12,\,50,\,52,\,54,\,56$ (right, (\ref{eq:uDboost2})).
Parameters are $N_f=1024$, $N_c = K^0_\Gamma = 64$ 
and $\tau=10^{-3}$.
}
\label{fig:2dhex_boost} 
\end{figure}

\begin{figure}
\center
\includegraphics[angle=-90,width=0.23\textwidth]{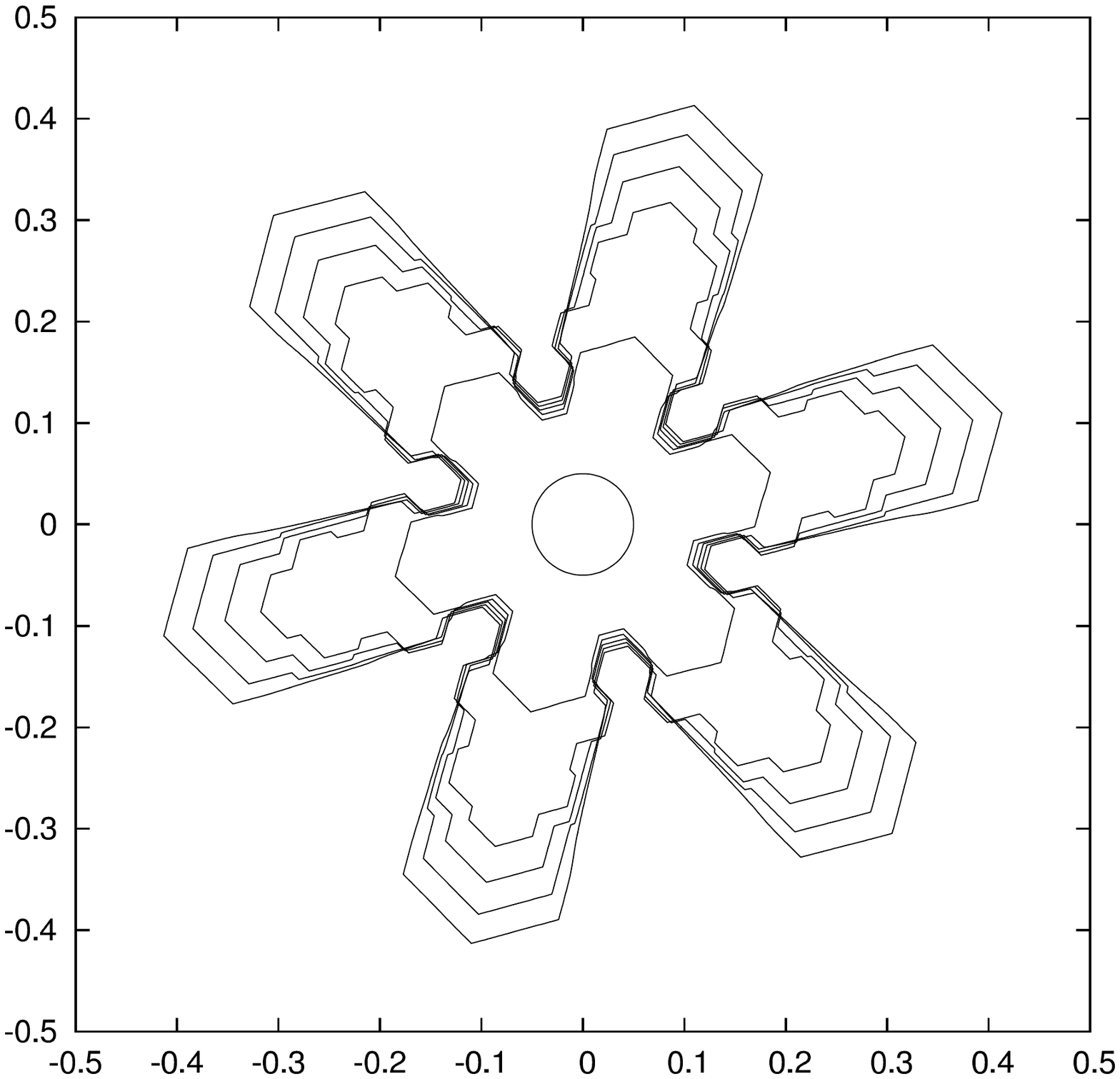}
\includegraphics[angle=-90,width=0.23\textwidth]{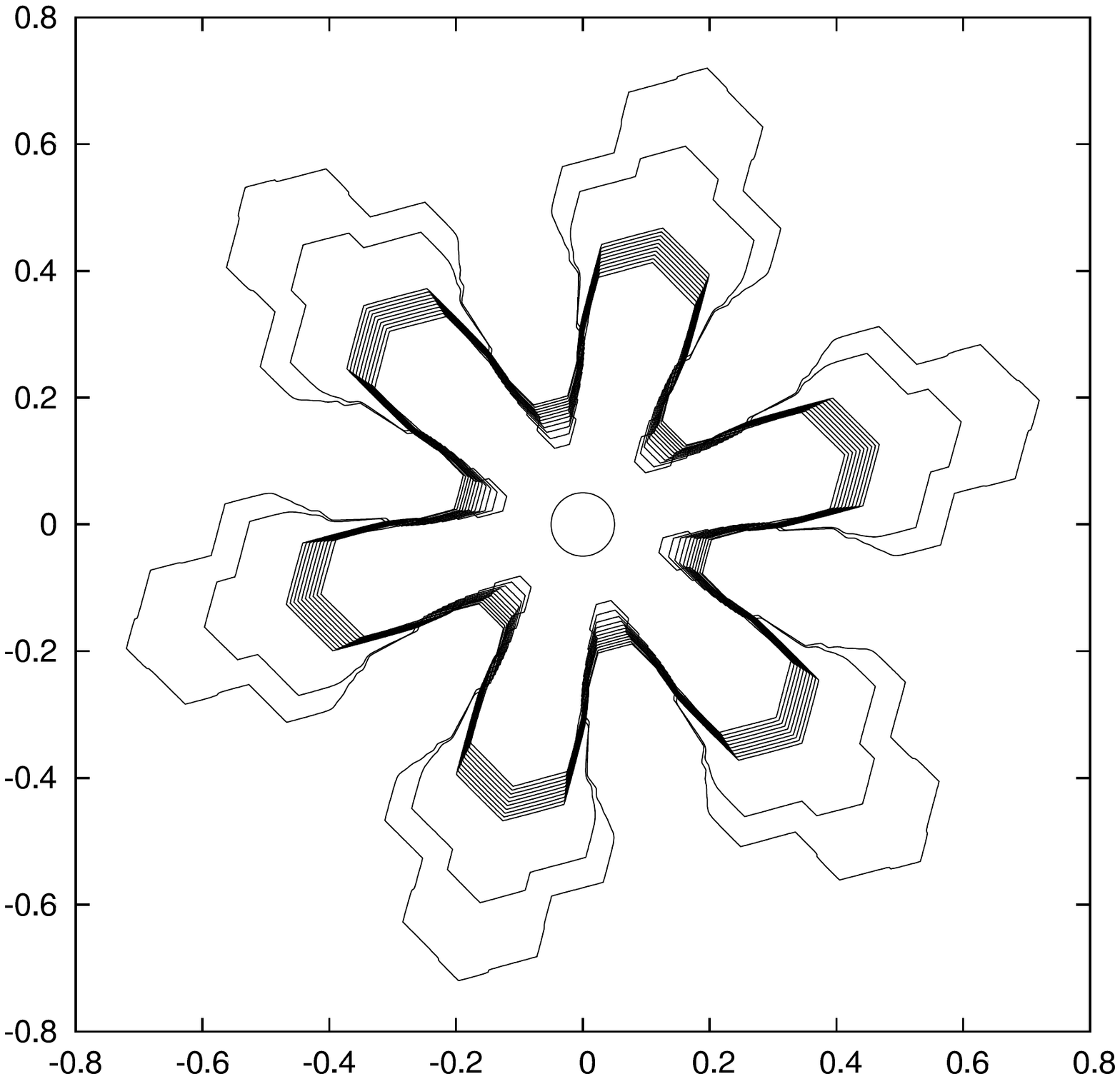}
\includegraphics[angle=-90,width=0.23\textwidth]{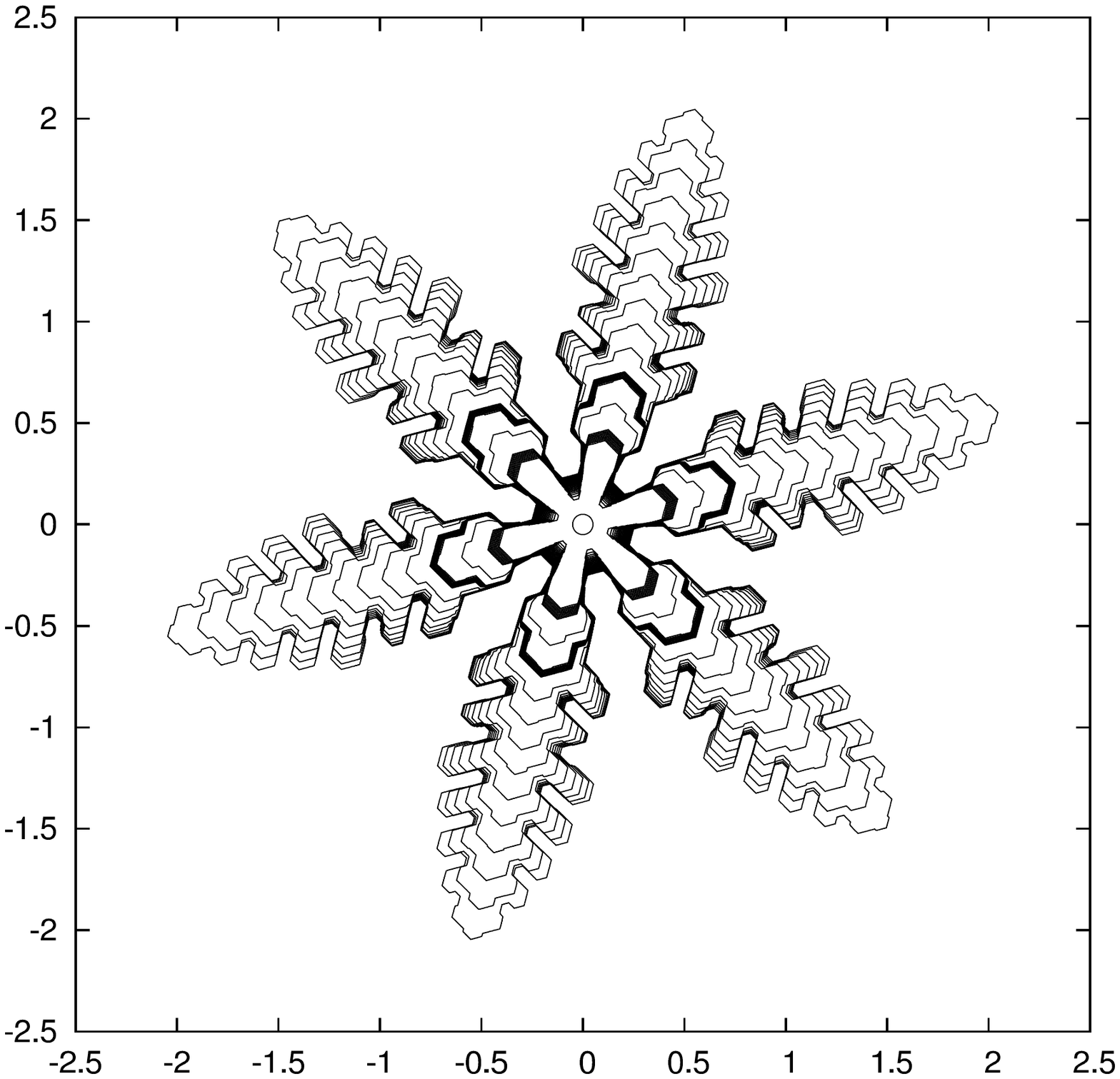}
\caption{($\Omega=(-4,4)^2$, $\gamma=\beta=\gamma_{hex}$)
$\Gamma^h(t)$ for $t=0,\,0.2,\ldots,1$ (left top), $t=0,\,1,\ldots,12$
(right top) and for $t=0,\,1,\ldots,30$ (bottom).
Parameters are $N_f=2048$, $N_c = K^0_\Gamma = 128$ 
and $\tau=2.5\times10^{-4}$.
}
\label{fig:2dms0} 
\end{figure}

We end this subsection with some computations, 
where we use a time dependent choice
for $\uD$. This models changing physical conditions. In
particular, in the first computation we set
\begin{equation} \label{eq:uDboost}
\uD(t) = \left\{\begin{array}{ll}
0.004  &t \in [0,50) \cup [60, 200] \,,\\
0.08  &t \in [50,60) \,;
\end{array}\right.
\end{equation}
while in the second computation we set
\begin{equation} \label{eq:uDboost2}
\uD(t) = \left\{\begin{array}{ll}
0.08  &t \in [0,12) \cup [50, 56]\,,\\
0.004 & t \in [12,50) \,;
\end{array}\right.
\end{equation}
see Figure~\ref{fig:2dhex_boost} for the results.
In the last computation we use a widely varying $\uD$. 
We set 
$\uD(t) = 0.2$ for $t \in[0,0.2)$, then
$\uD(t) = 0.4$ till $t=0.3$, then
$\uD(t) = 0.08$ till $t=1$, then
$\uD(t) = 0.004$ till $t=10$, then
$\uD(t) = 0.08$ till $t=12$, then
$\uD(t) = 0.004$ till $t=20$, and then
$\uD(t) = 0.08$ until the end,
see Figure~\ref{fig:2dms0} for the results.

\subsection{Snow crystal simulations in three space dimensions}

\begin{figure}
\center
\includegraphics[angle=-90,totalheight=4cm]{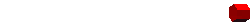} \qquad
\includegraphics[angle=-90,totalheight=4cm]{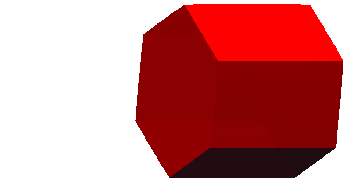} \qquad
\includegraphics[angle=-90,totalheight=4cm]{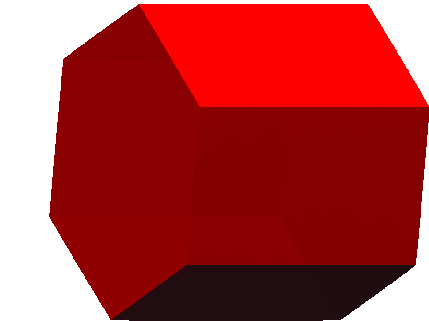}
\includegraphics[angle=-90,totalheight=3cm]{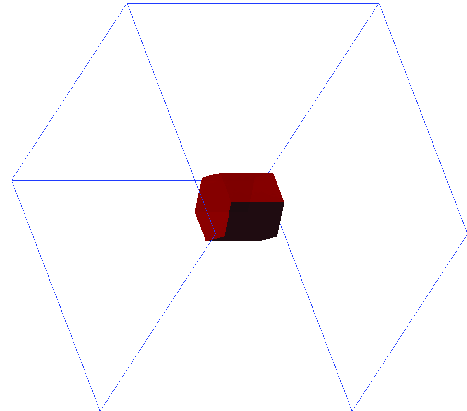} 
\caption{($\Omega=(-8,8)^3$, $\gamma=\beta=\gamma^{\rm TB}_{hex}$ 
with $\gamma_{\rm TB} = 0.95$)
$\Gamma^h(t)$ for $t=0,\,0.1,\,0.2$; and $\Gamma^h(0.2)$ within $\Omega$.
Parameters are $N_f=512$, $N_c=32$, $K^0_\Gamma = 1538$ and 
$\tau=10^{-4}$.}
\label{fig:prupp095} 
\end{figure}

Also in three space dimensions we use the physically relevant
parameters introduced in Section~\ref{sec2}, see in particular
(\ref{param}), and choose, if not stated otherwise,
the three-dimensional variant of
$\gamma_{hex}$, see (\ref{eq:L44}), with $\epsilon =0.01$ and
$\theta_0=\frac{\pi}{12}$. In all computations with the exception of
Figure~\ref{fig:prupp095} the initial crystal seed was
spherical with radius $0.05$. Similarly to our computations 
in two space dimensions, we
observe in our three dimensional numerical computations 
that the surface energy anisotropy
is important in order to obtain facetted growth. If we do not choose
the surface energy strongly facetted, then we do not observe facetted growth
of the crystal.

One issue in three dimensions is to understand how the parameter $\beta$
leads to either horizontal flat growth or to columnar vertical growth, which 
may yield e.g.\ solid prisms or needles, respectively. 
First of all, we attempt to compute a
self-similar hexagonal evolution, i.e.\ 
a crystal where the basal and prismal facets
grow with the same velocity. This is motivated by a theoretical
result in \cite{GigaR04}, in which the existence of self-similar
evolutions of crystals, where the Wulff shape is a
cylinder, was shown. 
We choose $\rho=\alpha=1$, $\uD=21$, $\gamma=\beta=\gamma^{\rm TB}_{hex}$ as in
(\ref{eq:L44gTB}), vary the ratio $\gamma_{\rm TB} =
\gamma^{\rm B}/\gamma^{\rm P}$ and observe that,
upon starting the evolution with $\Gamma(0)$ being a scaled Wulff shape, 
for $\gamma_{\rm TB} \approx 0.95$ the evolution is 
self-similar up to discretization errors. See Figure~\ref{fig:prupp095}
for a computation with $\gamma_{\rm TB} = 0.95$.

\begin{figure}
\center
\includegraphics[angle=-90,totalheight=3cm]{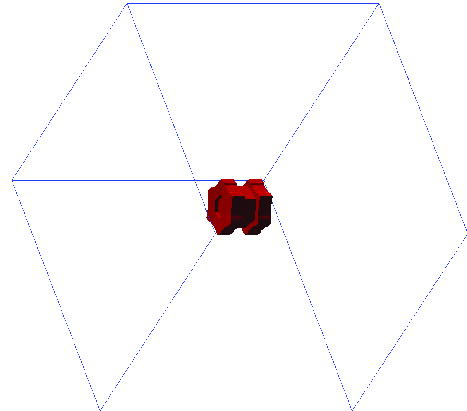}
\qquad
\includegraphics[angle=-90,totalheight=3cm]{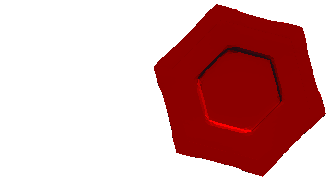}
\qquad
\includegraphics[angle=-90,totalheight=3cm]{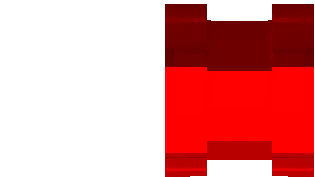}
\caption{($\Omega=(-4,4)^3$, $\uD = 0.004$, $\gamma=\gamma_{hex}$, $\beta = 1$)
$\Gamma^h(50)$.
Parameters are $N_f=128$, $N_c=16$, $K^0_\Gamma = 98$ 
and $\tau=10^{-1}$.}
\label{fig:20} 
\end{figure}

For the remainder of the computations we fix $\gamma_{\rm TB} = 1$, i.e.\ we
choose $\gamma = \gamma_{hex}$ as in (\ref{eq:L44}),
and use the physically relevant parameters in (\ref{param}).
For the first such computation we set $\uD=0.004$ and $\beta=1$, 
see Figure~\ref{fig:20}. We can clearly see
that the facets of the growing crystal are aligned with the Wulff
shape of $\gamma$. We also note that facet breaking occurs both
in the prismal and in the basal directions. In this context we refer
to \cite{GondaY82}, where similar facet breaking was
observed in experiments.

\begin{figure}
\center
\includegraphics[angle=-90,totalheight=2.5cm]{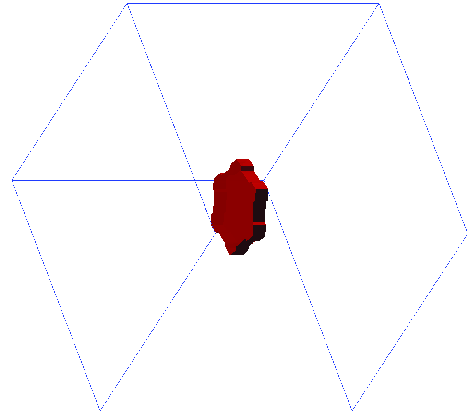}
\quad
\includegraphics[angle=-90,totalheight=2.5cm]{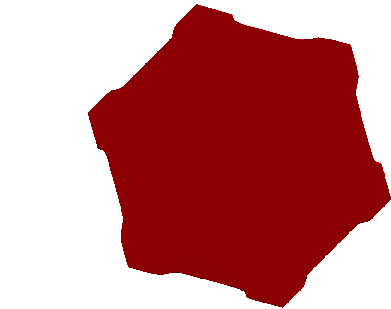}
\quad
\includegraphics[angle=-90,totalheight=2.5cm]{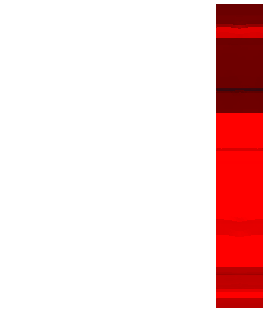}
\caption{($\Omega=(-4,4)^3$, $\uD = 0.004$, $\gamma=\gamma_{hex}$, 
$\beta = \beta_{\rm flat,2}$)
$\Gamma^h(50)$.
Parameters are $N_f=128$, $N_c=16$, $K^0_\Gamma = 98$ 
and $\tau=10^{-1}$.}
\label{fig:21} 
\end{figure}%

\begin{figure}
\center
\includegraphics[angle=-90,totalheight=3.5cm]{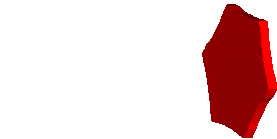}\quad
\includegraphics[angle=-90,totalheight=3.5cm]{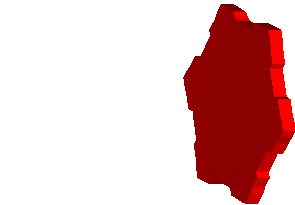}\quad
\includegraphics[angle=-90,totalheight=3.5cm]{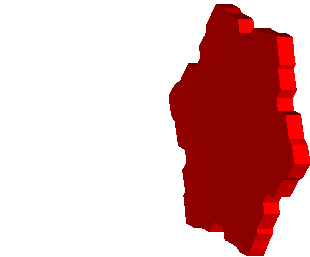}\quad
\includegraphics[angle=-90,totalheight=3.5cm]{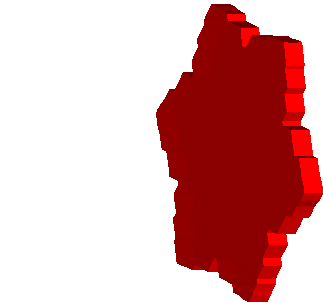}
\caption{($\Omega=(-8,8)^3$, $\uD = 0.004$, $\gamma=\gamma_{hex}$, 
$\beta = \beta_{\rm flat,3}$)
$\Gamma^h(t)$ for $t=50,\,100,\,150,\,200$.
Parameters are $N_f=256$, $N_c=32$, $K^0_\Gamma = 98$ 
and $\tau=10^{-1}$.}
\label{fig:23long} 
\end{figure}%
It is well known that the condensation coefficient $\beta$ varies
strongly for different orientations, depending on the meteorological
environment. 
In particular, the value of $\beta$ can differ quite drastically
between directions which correspond to basal facet normals and ones which
correspond to prismal facet normals, see \cite{Libbrecht05}. 
We hence perform
different numerical computations for the condensation coefficients
$\beta_{\rm flat}$ and $\beta_{\rm tall}$ defined in 
(\ref{eq:betaflat}) and (\ref{eq:betatall}).
 
We begin with a repeat of the simulation in Figure~\ref{fig:20}, but now choose
as kinetic coefficient $\beta = \beta_{\rm flat,2}$ 
and $\beta = \beta_{\rm flat,3}$; see 
Figures~\ref{fig:21} and \ref{fig:23long}.
In comparison to the evolution in Figure~\ref{fig:20} one observes
that the smaller condensation coefficient in basal directions leads to
flat crystals. This is related to shapes in the Nakaya diagram for
temperature between $0^\circ$C and $-3^\circ$C and between
$-10^\circ$C and $-22^\circ$C.

\begin{figure}
\center
\includegraphics[angle=-90,totalheight=3cm]{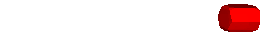} \quad
\includegraphics[angle=-90,totalheight=3cm]{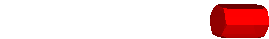} \quad
\includegraphics[angle=-90,totalheight=3cm]{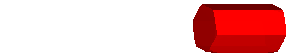} \quad
\includegraphics[angle=-90,totalheight=3cm]{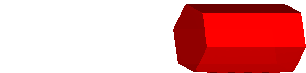} \quad
\includegraphics[angle=-90,totalheight=3cm]{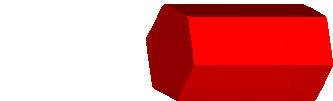} \quad
\includegraphics[angle=-90,totalheight=3cm]{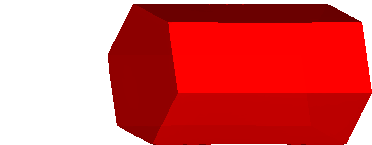} \quad
\includegraphics[angle=-90,totalheight=3cm]{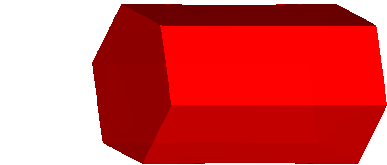} \quad
\includegraphics[angle=-90,totalheight=3cm]{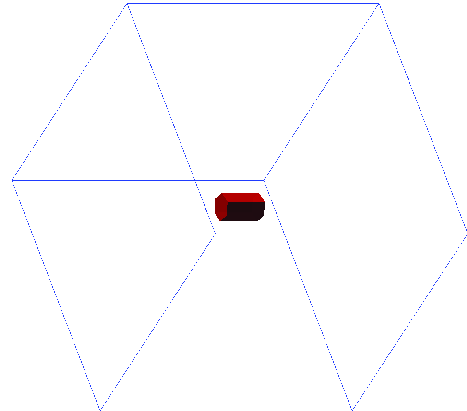} 
\caption{($\Omega=(-4,4)^3$, $\uD = 0.002$, $\gamma=\gamma_{hex}$,
$\beta = \beta_{\rm tall,1}$)
$\Gamma^h(t)$ for $t=1,\,2,\,5,\,10,\,20,
\,40,\,50$; 
and $\Gamma^h(50)$ within $\Omega$.
Parameters are $N_f=128$, $N_c=16$, $K^0_\Gamma = 98$ and $\tau=10^{-1}$.}
\label{fig:solidprism} 
\end{figure}

A computation with a supersaturation $\uD=0.002$ and $\beta =\beta_{\rm
  tall,1}$ can be seen in Figure~\ref{fig:solidprism}. In this case the
condensation coefficient is larger in the basal direction and we
obtain a solid prism, which can be found in the Nakaya diagram at
temperatures between $-5^\circ$C and $-10^\circ$C and at low
supersaturations. 

\begin{figure}
\center
\includegraphics[angle=-90,totalheight=3cm]{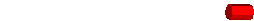} \quad
\includegraphics[angle=-90,totalheight=3cm]{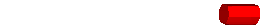} \quad
\includegraphics[angle=-90,totalheight=3cm]{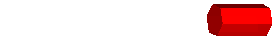} \quad
\includegraphics[angle=-90,totalheight=3cm]{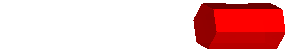} \quad
\includegraphics[angle=-90,totalheight=3cm]{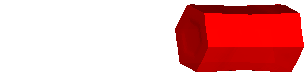} \quad
\includegraphics[angle=-90,totalheight=3cm]{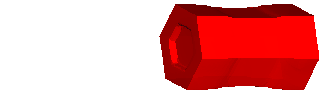} \quad
\includegraphics[angle=-90,totalheight=3cm]{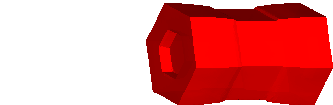} \quad
\includegraphics[angle=-90,totalheight=3cm]{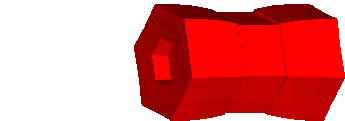} \quad
\includegraphics[angle=-90,totalheight=3cm]{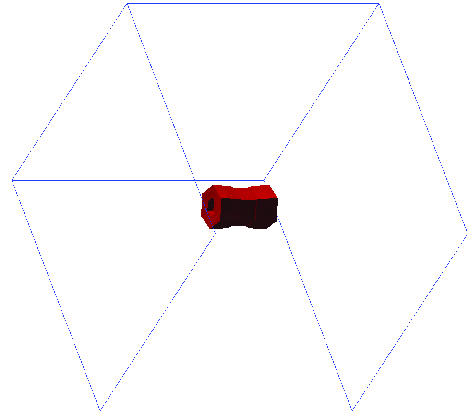} 
\caption{($\Omega=(-4,4)^3$, $\uD = 0.004$, $\gamma=\gamma_{hex}$,
$\beta = \beta_{\rm tall,1}$)
$\Gamma^h(t)$ for $t=1,\,2,\,5,\,10,\,20,\,30,\,40,\,50$; 
and $\Gamma^h(50)$ within $\Omega$.
Parameters are $N_f=128$, $N_c=16$, $K^0_\Gamma = 98$ and $\tau=10^{-1}$.}
\label{fig:hollowcolumn} 
\end{figure}

\begin{figure}
\center
\hspace*{-1cm}
\includegraphics[angle=-0,totalheight=3.5cm]{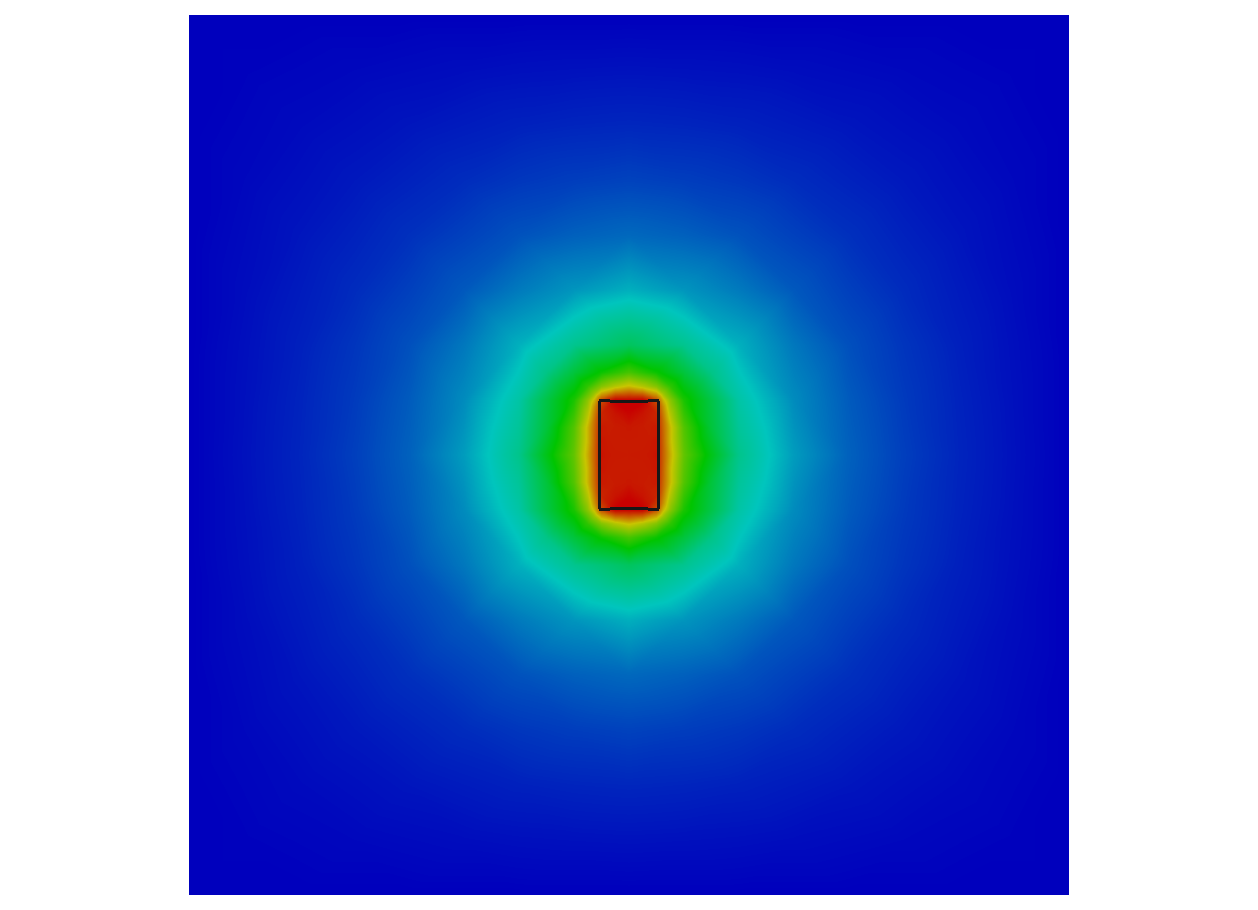}
\hspace*{-0.5cm}
\includegraphics[angle=-0,totalheight=3.5cm]{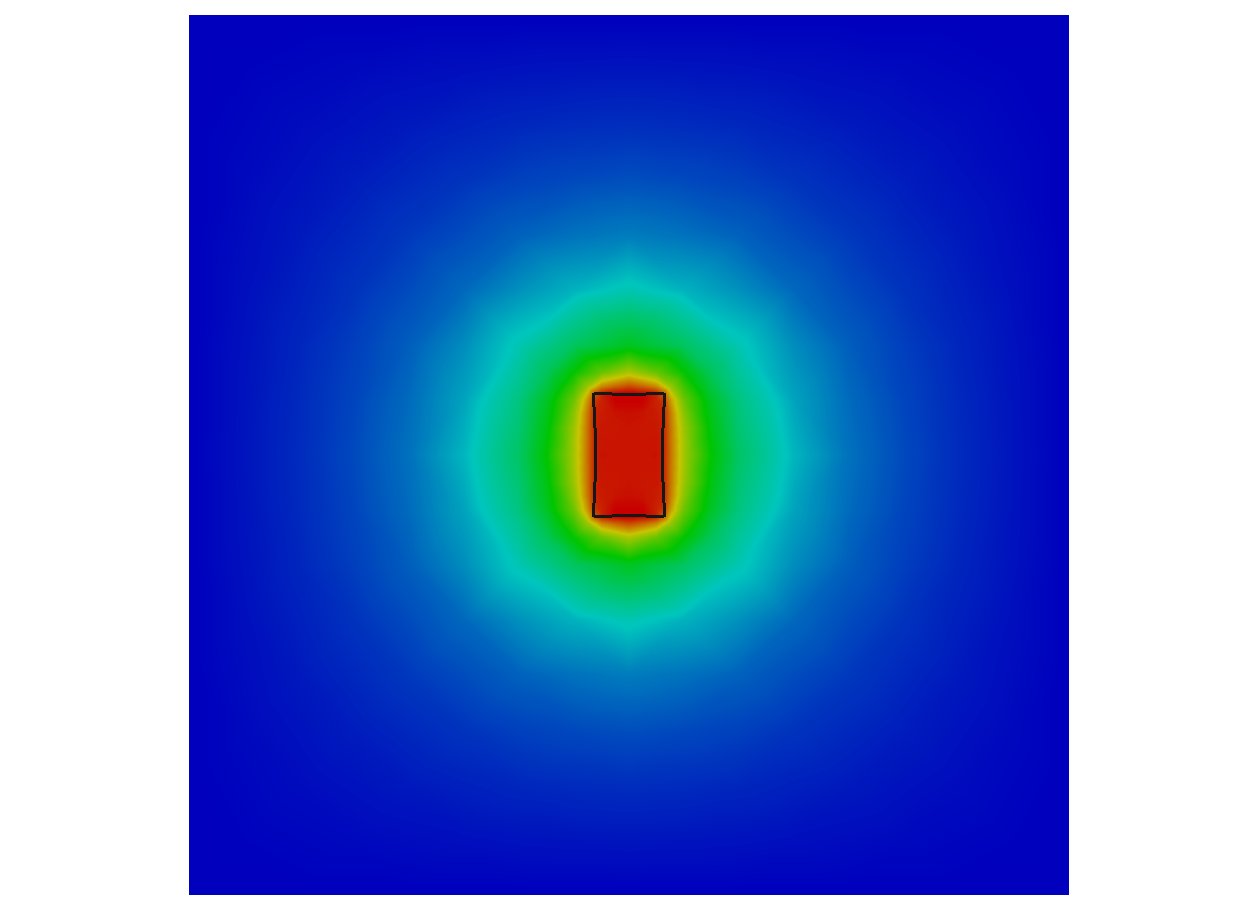}
\hspace*{-1cm}
\includegraphics[angle=-0,totalheight=3.5cm]{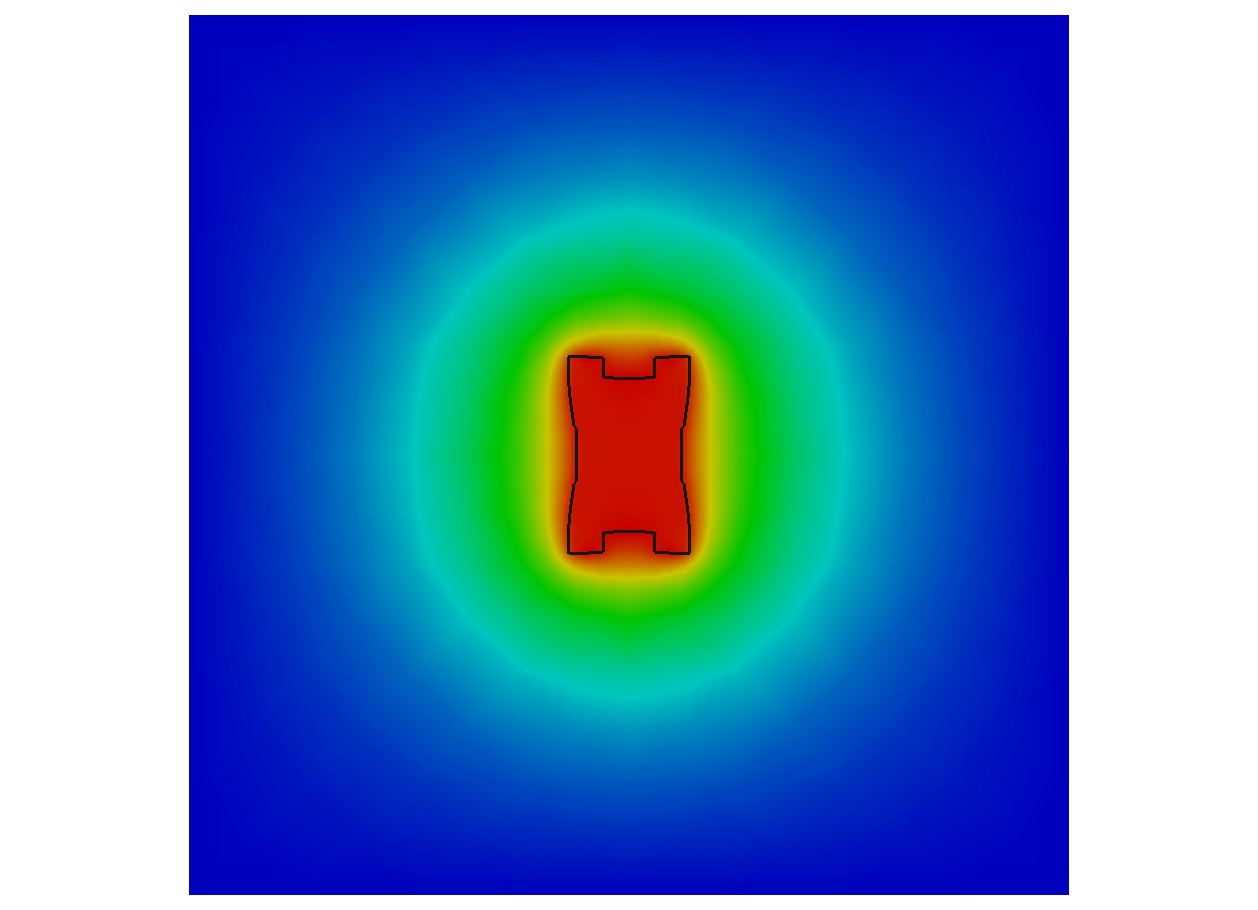}
\qquad
\caption{($\Omega=(-4,4)^3$, $\uD = 0.004$, $\gamma=\gamma_{hex}$,
$\beta = \beta_{\rm tall,1}$)
$\Gamma^h(t) \cap \{\vec x : x_1 = 0\}$ 
and $u^h(t)\!\mid_{x_1=0}$ for $t=15,\, 20,\, 50$.
The colours for $u^h$ vary between red for
$u^h = -1.12\times10^{-4}$ and blue for $u^h = 4\times10^{-3}$.
Parameters are $N_f=128$, $N_c=16$, $K^0_\Gamma = 98$ 
and $\tau=10^{-1}$.}
\label{fig:24temppng} 
\end{figure}

At higher supersaturations $\uD=0.004$ we obtain for 
$\beta=\beta_{\rm tall,1}$ the results shown in Figure~\ref{fig:hollowcolumn}.
We also give some plots of the rescaled water vapour density in
Figure~\ref{fig:24temppng}. 
We observe Berg's effect \cite{Berg38}, which states that the concentration
is largest at the edges and decreases towards the
centre of the facet. It is believed that facet breaking occurs when
the concentration becomes too non-uniform on the facets
\cite{GigaR03}. In Figure~\ref{fig:hollowcolumn} we observe facet
breaking for the basal and prismal directions, although the breaking
predominantly occurs on the basal facets. 

\begin{figure}
\center
\includegraphics[angle=-90,totalheight=5cm]{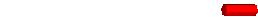} \quad
\includegraphics[angle=-90,totalheight=5cm]{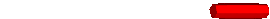} \quad
\includegraphics[angle=-90,totalheight=5cm]{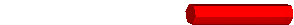} \quad
\includegraphics[angle=-90,totalheight=5cm]{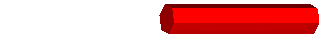} \quad
\includegraphics[angle=-90,totalheight=5cm]{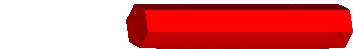} \quad
\includegraphics[angle=-90,totalheight=5cm]{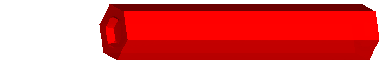} \quad
\includegraphics[angle=-90,totalheight=5cm]{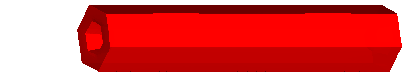} \quad
\includegraphics[angle=-90,totalheight=5cm]{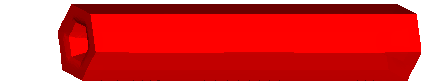} \quad
\includegraphics[angle=-90,totalheight=3cm]{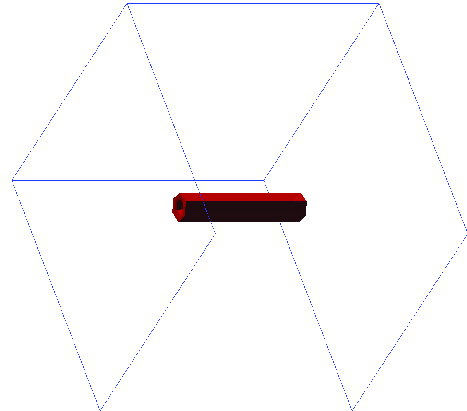}
\caption{($\Omega=(-4,4)^3$, $\uD = 0.008$, $\gamma=\gamma_{hex}$,
$\beta = \beta_{\rm tall,2}$)
$\Gamma^h(t)$ for $t=1,\,2,\,5,\,10,\,20,\,30,\,40,\,50$; 
and $\Gamma^h(50)$ within $\Omega$.
Parameters are $N_f=128$, $N_c=16$, $K^0_\Gamma = 98$ and $\tau=10^{-1}$.}
\label{fig:25} 
\end{figure}%

\begin{figure}
\center
\includegraphics[angle=-90,totalheight=6cm]{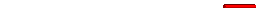} \quad
\includegraphics[angle=-90,totalheight=6cm]{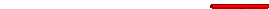} \quad
\includegraphics[angle=-90,totalheight=6cm]{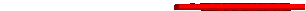} \quad
\includegraphics[angle=-90,totalheight=6cm]{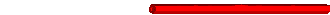} \quad
\includegraphics[angle=-90,totalheight=6cm]{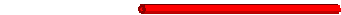} \quad
\qquad
\includegraphics[angle=-90,totalheight=6cm]{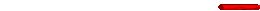} \quad
\includegraphics[angle=-90,totalheight=6cm]{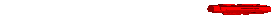} \quad
\includegraphics[angle=-90,totalheight=6cm]{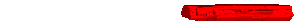} \quad
\includegraphics[angle=-90,totalheight=6cm]{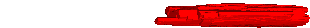} \quad
\includegraphics[angle=-90,totalheight=6cm]{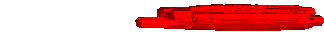} 
\includegraphics[angle=-90,totalheight=3cm]{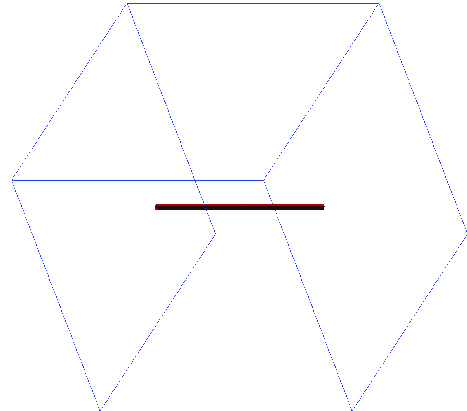}
\qquad\qquad
\includegraphics[angle=-90,totalheight=3cm]{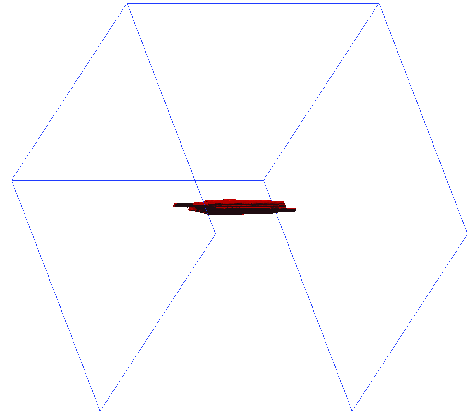}
\caption{($\Omega=(-8,8)^3$, $\uD = 0.004$, $\gamma=\gamma_{hex}$, 
$\beta = \beta_{\rm tall,3}$)
$\Gamma^h(t)$ for $t=5,\,10,\,30,\,50,\,60$; 
and $\Gamma^h(60)$ within $\Omega$.
Parameters are $N_f=512$, $N_c=32$, $K^0_\Gamma = 98$ and
$\tau=10^{-2}$ (left), and
$N_f=256$, $N_c=32$, $K^0_\Gamma = 98$ and $\tau=10^{-1}$ (right).}
\label{fig:27c004} 
\end{figure}

Choosing the condensation coefficient even larger in the basal
directions leads to Figure~\ref{fig:25}. We observe hollow columns
as in the Nakaya diagram between $-5^\circ$C and $-10^\circ$C at low,
but not too low, supersaturations. Increasing the condensation
coefficient in the basal directions even further, i.e.\ choosing
$\beta=\beta_{\rm tall,3}$, leads to the evolution depicted on the left of
Figure~\ref{fig:27c004}. 
On the right we also display a computation on a coarser grid. Both results in
Figure~\ref{fig:27c004} lead to needle growth, which also
appears in the Nakaya diagram. We remark that the shape on the right of
Figure~\ref{fig:27c004} is caused by numerical noise and
rounding errors. However, the same
effect, on even the most refined meshes, can be achieved by adding random
fluctuations to the model. In real life such
fluctuations and changes in physical parameters are experienced by the growing
snow crystal, as it moves through the atmosphere towards the earth.

\begin{figure}
\center
\includegraphics[angle=-90,totalheight=3cm]{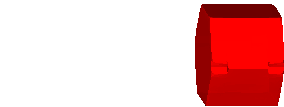} \quad
\includegraphics[angle=-90,totalheight=3cm]{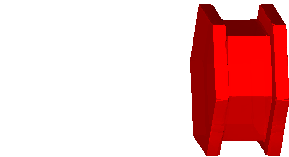} \quad
\includegraphics[angle=-90,totalheight=3cm]{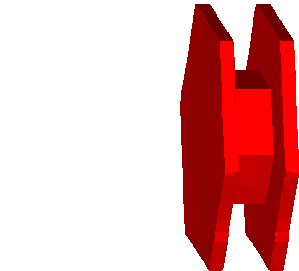} \quad
\includegraphics[angle=-90,totalheight=3cm]{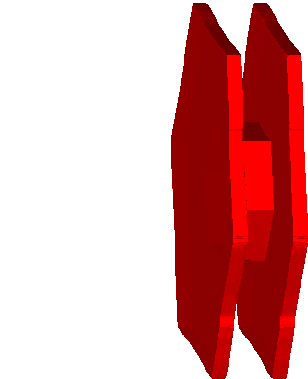} \quad
\includegraphics[angle=-90,totalheight=3cm]{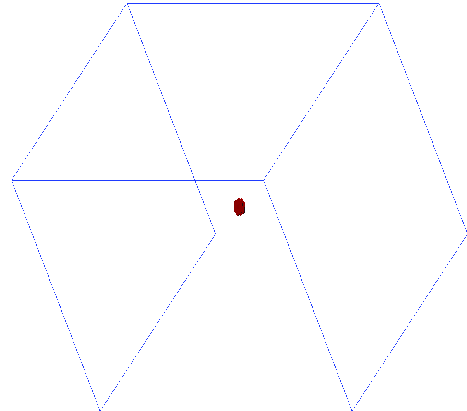}
\caption{($\Omega=(-4,4)^3$, $\uD = 0.02$, $\gamma=\gamma_{hex}$,
$\beta = \beta_{\rm flat,3}$)
$\Gamma^h(t)$ for $t=0.05,\,0.1,\,0.2,\,0.3$; 
and $\Gamma^h(0.3)$ within $\Omega$.
Parameters are $N_f=512$, $N_c=32$, $K^0_\Gamma = 1538$ 
and $\tau=5\times10^{-4}$.}
\label{fig:c02f6} 
\end{figure}%

A numerical simulation with supersaturation $\uD=0.02$ with
$\beta=\beta_{\rm flat,3}$ is displayed in Figure~\ref{fig:c02f6}. In
this case capped columns appear, which can also be observed in nature;
see \cite{Libbrecht05,Libbrecht06}.  

\begin{figure}
\center
\includegraphics[angle=-90,totalheight=5cm]{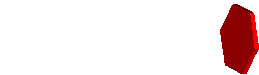} 
\includegraphics[angle=-90,totalheight=5cm]{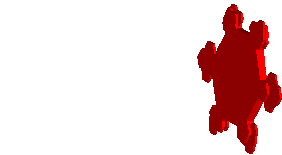} 
\includegraphics[angle=-90,totalheight=5cm]{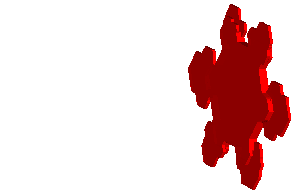} 
\includegraphics[angle=-90,totalheight=5cm]{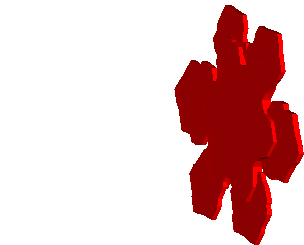} 
\includegraphics[angle=-90,totalheight=5cm]{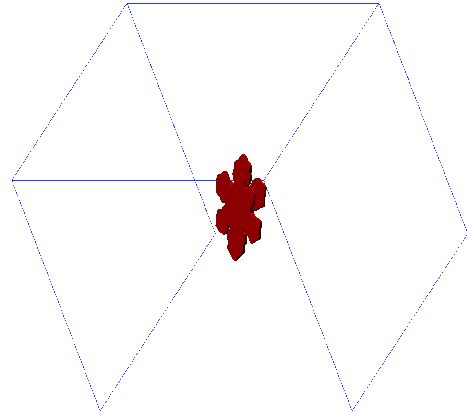}
\caption{($\Omega=(-8,8)^3$, $\uD$ as in (\ref{eq:L44boost}), 
$\gamma=\gamma_{hex}$,
$\beta = \beta_{\rm flat,3}$)
$\Gamma^h(t)$ for $t=15,\,20,\,30,
\,50$;
and $\Gamma^h(50)$ within $\Omega$.
Parameters are $N_f=512$, $N_c=32$, $K^0_\Gamma = 98$ and 
$\tau=2\times10^{-2}$.}
\label{fig:L44_c004_f6_boostlarge} 
\end{figure}

We end this subsection with computations, 
where we use a time dependent choice
for $\uD$. In particular, we set
\begin{equation} \label{eq:L44boost}
\uD(t) = \left\{\begin{array}{ll}
0.004 & t \in [0,15) \cup [18, 50] \,,\\
0.024 & t \in [15,18) \,.
\end{array}\right.
\end{equation}
See Figure~\ref{fig:L44_c004_f6_boostlarge} for the results.
First a plate forms and then, due to the fact that the supersaturation
increases, the plate becomes unstable and 
new plate-like shapes grow at the corners of the plate.

\begin{figure}
\center
\includegraphics[angle=-90,totalheight=3cm]{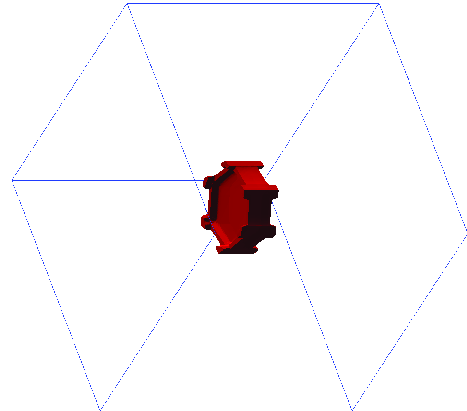}
\quad
\includegraphics[angle=-90,totalheight=3cm]{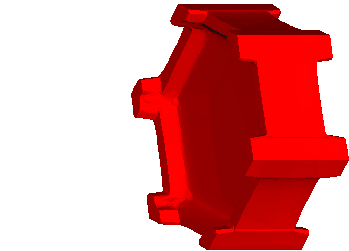}
\qquad\qquad
\includegraphics[angle=-90,totalheight=3cm]{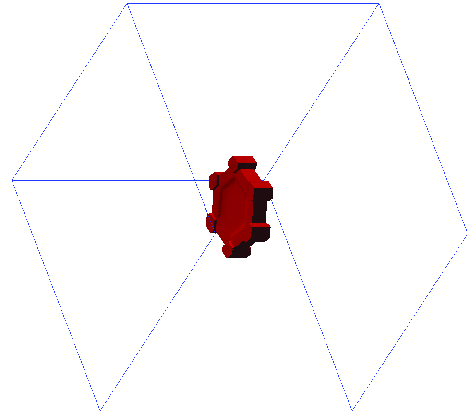}
\quad
\includegraphics[angle=-90,totalheight=3cm]{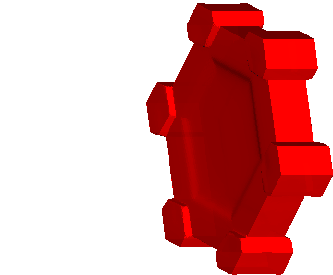}
\caption{($\Omega=(-4,4)^3$, $\uD = 0.004$, $\gamma=\gamma_{hex}$, 
$\beta$ as in (\ref{eq:beta_sw0}) (top), and as in (\ref{fig:22_switches}) 
(bottom)) 
$\Gamma^h(50)$. Parameters are $N_f=128$, $N_c=16$, $K^0_\Gamma = 98$ 
and $\tau=10^{-1}$.}
\label{fig:22_switches} 
\end{figure}

Finally, we perform two simulations, where we vary $\beta$ in time. In the first
such example, we choose
\begin{equation} \label{eq:beta_sw0}
\beta(\vec p) = 
\left\{\begin{array}{ll}
\beta_{\rm flat,3}(\vec p) & t \in [0,30)\,, \\
\beta_{\rm tall,3}(\vec p) & t \in [30,50] \,.
\end{array}\right.
\end{equation}
In a second example, we choose
\begin{equation} \label{eq:beta_sw1}
\beta(\vec p) = 
\left\{\begin{array}{ll}
\beta_{\rm flat,3}(\vec p) & t \in [0,20)\,, \\
\beta_{\rm flat,1}(\vec p) & t \in [20,50] \,.
\end{array}\right.
\end{equation}
Results for these choices of $\beta$ and for $\uD = 0.004$ can be seen in
Figure~\ref{fig:22_switches}.
We observe scrolls on plates, a shape that is also called 
plates with scrolls at ends, which also appear
in the Magono--Lee classification of natural snow crystals
\cite{MagonoL66}, see also 
\cite[p.~46]{PruppacherK97} and \cite{Libbrecht06}. 

\section{Conclusions}\label{sec5}

We have demonstrated that an approach introduced by the authors in
\cite{ani3d,dendritic,crystal} provides a powerful computational
tool to investigate pattern formation in crystal growth in a
qualitative and quantitative way. The method makes it possible to
simulate facetted and dendritic growth simultaneously. We also observe
the instability of crystals leading to facet breaking. Many parameters
in models for crystal growth are not known. The presented numerical
method in combination with a comparison to experiments can make it
possible to estimate the relative sizes of parameters. In particular,
by varying the condensation coefficient we were able to observe either
plate-like growth or columnar growth. 

Let us finally summarise the results.
\begin{itemize}
\item Surface energy effects taking anisotropy into account have been
included in the model and, despite their small size, they turned out to
totally change the character of the interfacial dynamics. In our
computations anisotropic
surface energy is required in the interfacial dynamics to produce
facetted dendritic growth. 
\item The influence of the anisotropy in the condensation coefficient,
  at least at small supersaturations, is not sufficient for facetted
  growth.
\item For small supersaturations the influence of the velocity term in
  (\ref{eq:2}) is small in comparison with the curvature term. 
\item The velocity at the tip of growing crystals depends in a linear
  way on the supersaturation. 
\item Macroscopic models for crystal growth which are based on a
  diffusion equation in the gas phase, a mass balance on the vapour
  crystal interface and a modified Gibbs--Thomson law, taking attachment
  kinetics into account, are able to model a variety of phenomena in
  crystal growth, such as the appearance of {\it solid plates}, {\it
    solid prisms},  {\it hollow columns}, {\it needles}, {\it
    dendrites}, {\it capped columns} and {\it scrolls on plates}. 
\end{itemize}

\noindent
{\bf Acknowledgement.} The authors wish to express their deep thanks
to Prof.\ K. G.\ Libbrecht of the California Institute of Technology for
many fruitful discussions and for providing
Figure~\ref{fig:libbrecht}.






\begin{thebibliography}{36}
\expandafter\ifx\csname natexlab\endcsname\relax\def\natexlab#1{#1}\fi
\providecommand{\bibinfo}[2]{#2}
\ifx\xfnm\relax \def\xfnm[#1]{\unskip,\space#1}\fi
\bibitem[{Nakaya(1954)}]{Nakaya54}
\bibinfo{author}{U.~Nakaya}, \bibinfo{title}{Snow Crystals: Natural and
  Artificial}, \bibinfo{publisher}{University Press},
  \bibinfo{address}{Cambridge}, \bibinfo{year}{1954}.
\bibitem[{Libbrecht(2005)}]{Libbrecht05}
\bibinfo{author}{K.~G. Libbrecht},
\newblock \bibinfo{title}{The physics of snow crystals},
\newblock \bibinfo{journal}{Rep. Progr. Phys.} \bibinfo{volume}{68}
  (\bibinfo{year}{2005}) \bibinfo{pages}{855--895}.
\bibitem[{Roosen and Taylor(1991)}]{RoosenT91}
\bibinfo{author}{A.~Roosen}, \bibinfo{author}{J.~E. Taylor},
\newblock \bibinfo{title}{Simulation of crystal growth with facetted
  interfaces},
\newblock \bibinfo{journal}{Mater. Res. Soc. Symp. Proc.} \bibinfo{volume}{237}
  (\bibinfo{year}{1991}) \bibinfo{pages}{25--36}.
\bibitem[{Yokoyama(1993)}]{Yokoyama93}
\bibinfo{author}{E.~Yokoyama},
\newblock \bibinfo{title}{Formation of patterns during growth of snow
  crystals},
\newblock \bibinfo{journal}{J. Cryst. Growth} \bibinfo{volume}{128}
  (\bibinfo{year}{1993}) \bibinfo{pages}{251--257}.
\bibitem[{Schmidt(1996)}]{Schmidt96}
\bibinfo{author}{A.~Schmidt},
\newblock \bibinfo{title}{Computation of three dimensional dendrites with
  finite elements},
\newblock \bibinfo{journal}{J. Comput. Phys.} \bibinfo{volume}{195}
  (\bibinfo{year}{1996}) \bibinfo{pages}{293--312}.
\bibitem[{Barrett et~al.(2010)Barrett, Garcke, and N{\"u}rnberg}]{dendritic}
\bibinfo{author}{J.~W. Barrett}, \bibinfo{author}{H.~Garcke},
  \bibinfo{author}{R.~N{\"u}rnberg},
\newblock \bibinfo{title}{On stable parametric finite element methods for the
  {S}tefan problem and the {M}ullins--{S}ekerka problem with applications to
  dendritic growth},
\newblock \bibinfo{journal}{J. Comput. Phys.} \bibinfo{volume}{229}
  (\bibinfo{year}{2010}) \bibinfo{pages}{6270--6299}.
\bibitem[{Sethian and Strain(1992)}]{SethianS92}
\bibinfo{author}{J.~A. Sethian}, \bibinfo{author}{J.~Strain},
\newblock \bibinfo{title}{Crystal growth and dendritic solidification},
\newblock \bibinfo{journal}{J. Comput. Phys.} \bibinfo{volume}{98}
  (\bibinfo{year}{1992}) \bibinfo{pages}{231--253}.
\bibitem[{Kobayashi(1993)}]{Kobayashi93}
\bibinfo{author}{R.~Kobayashi},
\newblock \bibinfo{title}{Modeling and numerical simulations of dendritic
  crystal growth},
\newblock \bibinfo{journal}{Phys. D} \bibinfo{volume}{63}
  (\bibinfo{year}{1993}) \bibinfo{pages}{410--423}.
\bibitem[{Wheeler et~al.(1993)Wheeler, Murray, and Schaefer}]{WheelerMS93}
\bibinfo{author}{A.~A. Wheeler}, \bibinfo{author}{B.~T. Murray},
  \bibinfo{author}{R.~J. Schaefer},
\newblock \bibinfo{title}{Computation of dendrites using a phase field model},
\newblock \bibinfo{journal}{Phys. D} \bibinfo{volume}{66}
  (\bibinfo{year}{1993}) \bibinfo{pages}{243--262}.
\bibitem[{Karma and Rappel(1998)}]{KarmaR98}
\bibinfo{author}{A.~Karma}, \bibinfo{author}{W.-J. Rappel},
\newblock \bibinfo{title}{Quantitative phase-field modeling of dendritic growth
  in two and three dimensions},
\newblock \bibinfo{journal}{Phys. Rev. E} \bibinfo{volume}{57}
  (\bibinfo{year}{1998}) \bibinfo{pages}{4323--4349}.
\bibitem[{Debierre et~al.(2003)Debierre, Karma, Celestini, and
  Gu{\'e}rin}]{DebierKCG03}
\bibinfo{author}{J.-M. Debierre}, \bibinfo{author}{A.~Karma},
  \bibinfo{author}{F.~Celestini}, \bibinfo{author}{R.~Gu{\'e}rin},
\newblock \bibinfo{title}{Phase-field approach for faceted solidification},
\newblock \bibinfo{journal}{Phys. Rev. E} \bibinfo{volume}{68}
  (\bibinfo{year}{2003}) \bibinfo{pages}{041604--1--13}.
\bibitem[{Reiter(2005)}]{Reiter05}
\bibinfo{author}{C.~A. Reiter},
\newblock \bibinfo{title}{A local cellular model for snow crystal growth},
\newblock \bibinfo{journal}{Chaos Soliton. Fract.} \bibinfo{volume}{23}
  (\bibinfo{year}{2005}) \bibinfo{pages}{1111--1119}.
\bibitem[{Libbrecht(2008)}]{Libbrecht08}
\bibinfo{author}{K.~G. Libbrecht}, \bibinfo{title}{Physically derived rules for
  simulating faceted crystal growth using cellular automata},
  \bibinfo{year}{2008}. \bibinfo{note}{\url{http://arxiv.org/abs/0807.2616}}.
\bibitem[{Gravner and Griffeath(2009)}]{GravnerG09}
\bibinfo{author}{J.~Gravner}, \bibinfo{author}{D.~Griffeath},
\newblock \bibinfo{title}{Modeling snow-crystal growth: A three-dimensional
  mesoscopic approach},
\newblock \bibinfo{journal}{Phys. Rev. E} \bibinfo{volume}{79}
  (\bibinfo{year}{2009}) \bibinfo{pages}{011601--1--18}.
\bibitem[{Furukawa and Nada(1997)}]{FurukawaN97}
\bibinfo{author}{Y.~Furukawa}, \bibinfo{author}{H.~Nada},
\newblock \bibinfo{title}{Anisotropic surface melting of an ice crystal and its
  relationship to growth forms},
\newblock \bibinfo{journal}{J. Phys. Chem. B}  (\bibinfo{year}{1997})
  \bibinfo{pages}{6167--6170}.
\bibitem[{Barrett et~al.(2007)Barrett, Garcke, and N{\"u}rnberg}]{triplej}
\bibinfo{author}{J.~W. Barrett}, \bibinfo{author}{H.~Garcke},
  \bibinfo{author}{R.~N{\"u}rnberg},
\newblock \bibinfo{title}{A parametric finite element method for fourth order
  geometric evolution equations},
\newblock \bibinfo{journal}{J. Comput. Phys.} \bibinfo{volume}{222}
  (\bibinfo{year}{2007}) \bibinfo{pages}{441--462}.
\bibitem[{Barrett et~al.(2008{\natexlab{a}})Barrett, Garcke, and
  N{\"u}rnberg}]{gflows3d}
\bibinfo{author}{J.~W. Barrett}, \bibinfo{author}{H.~Garcke},
  \bibinfo{author}{R.~N{\"u}rnberg},
\newblock \bibinfo{title}{On the parametric finite element approximation of
  evolving hypersurfaces in {${\mathbb R}^3$}},
\newblock \bibinfo{journal}{J. Comput. Phys.} \bibinfo{volume}{227}
  (\bibinfo{year}{2008}{\natexlab{a}}) \bibinfo{pages}{4281--4307}.
\bibitem[{Barrett et~al.(2008{\natexlab{b}})Barrett, Garcke, and
  N{\"u}rnberg}]{triplejANI}
\bibinfo{author}{J.~W. Barrett}, \bibinfo{author}{H.~Garcke},
  \bibinfo{author}{R.~N{\"u}rnberg},
\newblock \bibinfo{title}{Numerical approximation of anisotropic geometric
  evolution equations in the plane},
\newblock \bibinfo{journal}{IMA J. Numer. Anal.} \bibinfo{volume}{28}
  (\bibinfo{year}{2008}{\natexlab{b}}) \bibinfo{pages}{292--330}.
\bibitem[{Barrett et~al.(2008{\natexlab{c}})Barrett, Garcke, and
  N{\"u}rnberg}]{ani3d}
\bibinfo{author}{J.~W. Barrett}, \bibinfo{author}{H.~Garcke},
  \bibinfo{author}{R.~N{\"u}rnberg},
\newblock \bibinfo{title}{A variational formulation of anisotropic geometric
  evolution equations in higher dimensions},
\newblock \bibinfo{journal}{Numer. Math.} \bibinfo{volume}{109}
  (\bibinfo{year}{2008}{\natexlab{c}}) \bibinfo{pages}{1--44}.
\bibitem[{Barrett et~al.(2010)Barrett, Garcke, and N{\"u}rnberg}]{ejam3d}
\bibinfo{author}{J.~W. Barrett}, \bibinfo{author}{H.~Garcke},
  \bibinfo{author}{R.~N{\"u}rnberg},
\newblock \bibinfo{title}{Finite element approximation of coupled surface and
  grain boundary motion with applications to thermal grooving and sintering},
\newblock \bibinfo{journal}{European J. Appl. Math.} \bibinfo{volume}{21}
  (\bibinfo{year}{2010}) \bibinfo{pages}{519--556}.
\bibitem[{Libbrecht et~al.(2002)Libbrecht, Crosby, and Swanson}]{LibbrechtCS02}
\bibinfo{author}{K.~G. Libbrecht}, \bibinfo{author}{T.~Crosby},
  \bibinfo{author}{M.~Swanson},
\newblock \bibinfo{title}{Electrically enhanced free dendrite growth in polar
  and non-polar systems},
\newblock \bibinfo{journal}{J. Cryst. Growth} \bibinfo{volume}{240}
  (\bibinfo{year}{2002}) \bibinfo{pages}{241--254}.
\bibitem[{Ben-Jacob(1993)}]{BenJacob93}
\bibinfo{author}{E.~Ben-Jacob},
\newblock \bibinfo{title}{From snowflake formation to growth of bacterial
  colonies. {P}art {I}. {D}iffusive patterning in azoic systems},
\newblock \bibinfo{journal}{Contemp. Phys.} \bibinfo{volume}{34}
  (\bibinfo{year}{1993}) \bibinfo{pages}{247--273}.
\bibitem[{Cahn and Hoffman(1974)}]{CahnH74}
\bibinfo{author}{J.~W. Cahn}, \bibinfo{author}{D.~W. Hoffman},
\newblock \bibinfo{title}{A vector thermodynamics for anisotropic surfaces --
  {II}. {C}urved and faceted surfaces},
\newblock \bibinfo{journal}{Acta Metall.} \bibinfo{volume}{22}
  (\bibinfo{year}{1974}) \bibinfo{pages}{1205--1214}.
\bibitem[{Taylor et~al.(1992)Taylor, Cahn, and Handwerker}]{TaylorCH92}
\bibinfo{author}{J.~E. Taylor}, \bibinfo{author}{J.~W. Cahn},
  \bibinfo{author}{C.~A. Handwerker},
\newblock \bibinfo{title}{Geometric models of crystal growth},
\newblock \bibinfo{journal}{Acta Metall. Mater.} \bibinfo{volume}{40}
  (\bibinfo{year}{1992}) \bibinfo{pages}{1443--1474}.
\bibitem[{Davis(2001)}]{Davis01}
\bibinfo{author}{S.~H. Davis}, \bibinfo{title}{Theory of Solidification},
  Cambridge Monographs on Mechanics, \bibinfo{publisher}{Cambridge University
  Press}, \bibinfo{address}{Cambridge}, \bibinfo{year}{2001}.
\bibitem[{Gonda and Yamazaki(1982)}]{GondaY82}
\bibinfo{author}{T.~Gonda}, \bibinfo{author}{T.~Yamazaki},
\newblock \bibinfo{title}{Morphological stability of polyhedral ice crystals
  growing from the vapor phase},
\newblock \bibinfo{journal}{J. Cryst. Growth} \bibinfo{volume}{60}
  (\bibinfo{year}{1982}) \bibinfo{pages}{259--263}.
\bibitem[{Barrett et~al.(2012)Barrett, Garcke, and N{\"u}rnberg}]{crystal}
\bibinfo{author}{J.~W. Barrett}, \bibinfo{author}{H.~Garcke},
  \bibinfo{author}{R.~N{\"u}rnberg}, \bibinfo{title}{Finite element
  approximation of one-sided {S}tefan problems with anisotropic, approximately
  crystalline, {G}ibbs--{T}homson law}, \bibinfo{year}{2012}.
  \bibinfo{note}{\url{http://arxiv.org/abs/1201.1802v1}}.
\bibitem[{Gurtin(1993)}]{Gurtin93}
\bibinfo{author}{M.~E. Gurtin}, \bibinfo{title}{Thermomechanics of Evolving
  Phase Boundaries in the Plane}, Oxford Mathematical Monographs,
  \bibinfo{publisher}{The Clarendon Press Oxford University Press},
  \bibinfo{address}{New York}, \bibinfo{year}{1993}.
\bibitem[{Pruppacher and Klett(1997)}]{PruppacherK97}
\bibinfo{author}{H.~R. Pruppacher}, \bibinfo{author}{J.~D. Klett},
  \bibinfo{title}{Microphysics of Clouds and Precipitation},
  \bibinfo{publisher}{Kluwer Acad. Publ.}, \bibinfo{address}{Dordrecht},
  \bibinfo{year}{1997}.
\bibitem[{Bonzel(2003)}]{Bonzel03}
\bibinfo{author}{H.~P. Bonzel},
\newblock \bibinfo{title}{3{D} equilibrium crystal shapes in the new light of
  {STM} and {AFM}},
\newblock \bibinfo{journal}{Phys. Rep.} \bibinfo{volume}{385}
  (\bibinfo{year}{2003}) \bibinfo{pages}{1--67}.
\bibitem[{Libbrecht(2012)}]{Libbrecht_private}
\bibinfo{author}{K.~G. Libbrecht}, \bibinfo{title}{(private communication)},
  \bibinfo{year}{2012}.
\bibitem[{Giga and Rybka(2004)}]{GigaR04}
\bibinfo{author}{Y.~Giga}, \bibinfo{author}{P.~Rybka},
\newblock \bibinfo{title}{Existence of self-similar evolution of crystals grown
  from supersaturated vapor},
\newblock \bibinfo{journal}{Interfaces Free Bound.} \bibinfo{volume}{6}
  (\bibinfo{year}{2004}) \bibinfo{pages}{405--421}.
\bibitem[{Berg(1938)}]{Berg38}
\bibinfo{author}{W.~F. Berg},
\newblock \bibinfo{title}{Crystal growth from solutions},
\newblock \bibinfo{journal}{Proc. Roy. Soc. London Ser. A}
  \bibinfo{volume}{164} (\bibinfo{year}{1938}) \bibinfo{pages}{79--95}.
\bibitem[{Giga and Rybka(2003)}]{GigaR03}
\bibinfo{author}{Y.~Giga}, \bibinfo{author}{P.~Rybka},
\newblock \bibinfo{title}{Berg's effect},
\newblock \bibinfo{journal}{Adv. Math. Sci. Appl.} \bibinfo{volume}{13}
  (\bibinfo{year}{2003}) \bibinfo{pages}{625--637}.
\bibitem[{Libbrecht(2006)}]{Libbrecht06}
\bibinfo{author}{K.~G. Libbrecht}, \bibinfo{title}{Field Guide to Snowflakes},
  \bibinfo{publisher}{Voyageur Press}, \bibinfo{year}{2006}.
\bibitem[{Magono and Lee(1966)}]{MagonoL66}
\bibinfo{author}{C.~Magono}, \bibinfo{author}{C.~W. Lee},
\newblock \bibinfo{title}{Meteorological classification of natural snow
  crystals},
\newblock \bibinfo{journal}{J. Fac. Sci., Hokkaido Univ., Ser. VII}
  \bibinfo{volume}{2} (\bibinfo{year}{1966}) \bibinfo{pages}{321--335}.

\end{thebibliography}







\end{document}